# Using Reports of Symptoms and Diagnosis on Social Media to Predict COVID-19 Case Counts in Mainland China: Observational Infoveillance Study

Cuihua Shen[1*], Anfan Chen[2*], Chen Luo[3], Jingwen Zhang[1], Bo Feng[1], Wang Liao[1]

1. University of California, Davis
2. University of Science and Technology of China
3. Tsinghua University
*   Equal contribution

**Abstract**

**Background**: COVID-19 has affected more than 200 countries and territories worldwide. It poses an extraordinary challenge for public health systems, because screening and surveillance capacity—especially during the beginning of the outbreak—is often severely limited, fueling the outbreak as many patients unknowingly infect others.

**Objective**: We present an effort to collect and analyze COVID-19 related posts on the popular Twitter-like social media site in China, Weibo. To our knowledge, this infoveillance study employs the largest, most comprehensive and fine-grained social media data to date to predict COVID-19 case counts in mainland China.

**Methods**: We built a Weibo user pool of 250 million, approximately half of the entire monthly active Weibo user population. Using a comprehensive list of 167 keywords, we retrieved and analyzed around 15 million COVID-19 related posts from our user pool, from November 1, 2019 to March 31, 2020. We developed a machine learning classifier to identify "sick posts," which are reports of one's own and other people's symptoms and diagnosis related to COVID-19. Using officially reported case counts as the outcome, we then estimated the Granger causality of sick posts and other COVID-19 posts on daily case counts. For a subset of geotagged posts (3.10% of all retrieved posts), we also ran separate predictive models for Hubei province, the epicenter of the initial outbreak, and the rest of mainland China.

**Results**: We found that reports of symptoms and diagnosis of COVID-19 significantly predicted daily case counts, up to 14 days ahead of official statistics. But other COVID-19 posts did not have similar predictive power. For the subset of geotagged posts, we found that the predictive pattern held true for both Hubei province and the rest of mainland China, regardless of unequal distribution of healthcare resources and outbreak timeline.

**Conclusions**: Public social media data can be usefully harnessed to predict infection cases and inform timely responses. Researchers and disease control agencies should pay close attention to the social media infosphere regarding COVID-19. On top of monitoring overall search and posting activities, leveraging machine learning approaches and theoretical understandings of information sharing behaviors to identify true disease signals is a promising approach to improve the effectiveness of infoveillance.

**Keywords**: COVID-19, SARS-CoV-2, novel coronavirus, social media, Weibo, China, disease surveillance, infoveillance, infodemiology

**Introduction**

Since the outbreak of COVID-19 in December, 2019, in Wuhan, Hubei Province, China [1, 2], the novel coronavirus has already affected more than 200 countries and territories worldwide. As of May 16, 2020, there were more than 4 million confirmed cases and over 300,000 deaths [3]. Amid many unknowns, severe lack of laboratory testing capacity, delays in case reports, variations in local COVID-19 responses, and uncoordinated communication pose tremendous challenges for monitoring the epidemic dynamics and developing policies and targeted interventions for resource allocation.

When conventional disease surveillance capacity is limited, publicly available social media and Internet data can play a crucial role in uncovering hidden dynamics of an emerging outbreak [4]. Research in digital disease surveillance, also referred to as infoveillance or infodemiology, has shown great promise that Internet data can be usefully employed to track the real-time development of public attention, sentiment and health [5-8]. Specifically, data based on Internet searches and social media activities could nowcast and forecast disease prevalence as a supplement to conventional surveillance methods for various infectious diseases [5-7, 9-14].

One of the most well-known examples of digital disease surveillance is the case of Google Flu Trends, which used real-time Google search terms to predict clinical incidence rates of influenza with great initial success [13, 14]. Social media data such as Twitter were also shown to be effective in predicting and tracking various epidemics, such as influenza [10, 12] and Zika [15], with varying degrees of success. Yet, digital surveillance data also present unique challenges. For example, after its release in 2008, Google Flu Trends became less accurate over time, consistently overestimating flu prevalence during 2011-2013. The prediction error was partially attributed to people's changing search behaviors as well as increased public attention to the epidemic itself, which fueled awareness-related search queries with little to do with disease incidence [7, 16]. Compared to aggregated search queries, user-generated social media data have the advantage of being more direct and granular, allowing researchers to mine specific content to reflect actual illness. Still, media attention to emerging diseases can fuel social media activities, resulting in a deluge of discussions that dilute true disease signals on actual infection cases, making predictions less accurate [12].

The unprecedented magnitude and transmission speed of COVID-19 brought about massive social media activities as people isolate in their homes to break the infection chains [17]. Massive social media data inevitably contain massive noise (e.g., public reactions and awareness of the disease), which can be counterproductive for disease forecasting. A few early infoveillance studies have tracked public discussion of COVID-19 and patient characteristics on Weibo, the most popular public social media site in China [18-21]. Two studies suggest that COVID-19 related Weibo posts and search queries can be used to predict disease prevalence [19, 22]. However, they relied upon coarse-grained social media data and/or query data based on a few keywords with a short time window at the onset of the outbreak [19, 22]. As such, these studies' predictive accuracy and result interpretability are limited by the same pitfalls of infoveillance studies mentioned above. There are many reasons to search and discuss COVID-19 on social media, especially as the disease received substantial media coverage and most of the country was under mandatory lockdown. To more accurately predict infection cases and inform a rapid response, it is therefore critical to use granular and specific social media data to identify reliable disease signals (i.e., sick posts reporting symptoms and diagnosis).

Here we present an infoveillance effort to collect and analyze COVID-19 related posts on Weibo, and to identify specific type of Weibo posts that can predict COVID-19 case counts in mainland China. To our knowledge, this study collects the largest, most comprehensive and granular social media data related to COVID-19 in the Chinese language, far exceeding the scale, granularity and time span of similar studies [19, 22]. We built a Weibo user pool of 250 million, approximately half of the active Weibo user population [23]. Using a comprehensive list of 167 keywords associated with COVID-19, we retrieved around 15 million social media posts from November $1^{st}$, 2019 to March $31^{st}$, 2020. With much increased data granularity, we developed a supervised machine-learning classifier to distinguish "sick posts," which are reports of own and others' symptoms or diagnosis, from other COVID-19 related posts that could dilute disease signals from the data stream. Using the officially reported case accounts as the outcome, we compared the predictive power of sick posts versus other COVID-19 posts. We show evidence that sick posts predicted China CDC's daily cases up to 14 days in advance, while other COVID-19 related posts have much weaker predictive power. For the subset of geotagged posts, we found that the predictive pattern held true for both Hubei province and the rest of mainland China. Our work demonstrates one viable way to identify disease signals through reports of symptoms or diagnosis, rather than relying upon general discussion of COVID-19. It achieves a high level of prediction efficacy without sacrificing ease of interpretation, bringing significant contributions to the literature of infoveillance.

**Methods**

**Data Collection**

Social media data used in this study were collected from a popular Chinese microblog platform, Weibo, which had over 516 million monthly active users at the end of 2019 [23]. Weibo is very similar to Twitter, the access to which is blocked in mainland China. Unlike Twitter, Weibo does not provide large-scale public API access to its database. Keywords-based advanced search of Weibo posts is allowed via its web interface; however, the output of such search is limited to 50 pages (or around 1,000 posts) as per Weibo policy, making large-scale public data access notoriously difficult.

To bypass these limitations, we employed a Weibo user pool originally built in 2018, which started from 5 million active Weibo users obtained in our previous research unrelated to COVID-19 [24, 25]. We then retrieved the initial 5 million users' followers and followees (2nd degree users), the followers and followees of the 2nd degree users (3rd degree users), and so forth, until no new users were found. This snowball process resulted in a pool of 250 million users (with bots filtered out), which are approximately 48.4% of all monthly active Weibo users in 2019 and similar to the population of Weibo users in terms of self-reported sex and age distribution [26] (see Figure 1).

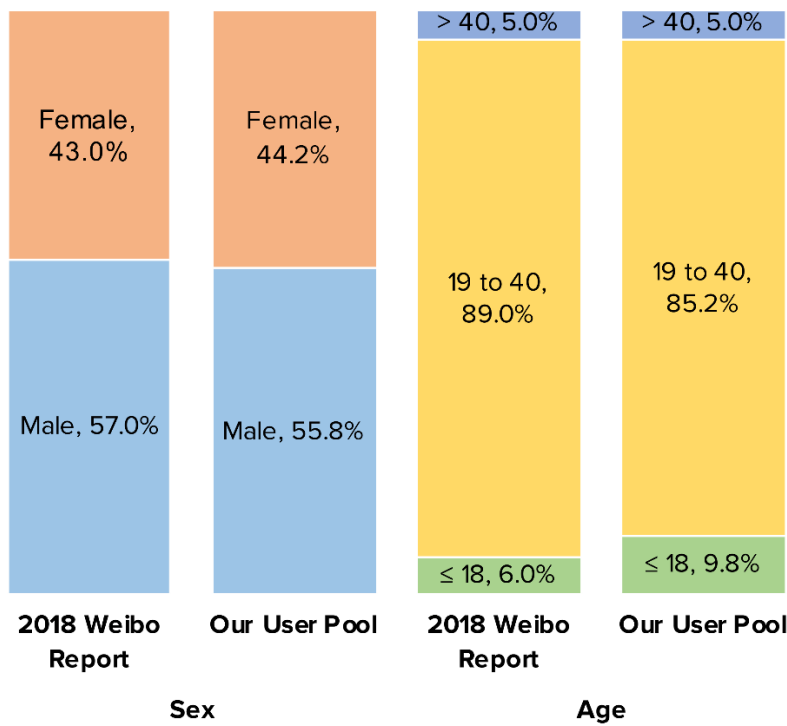

*Figure 1* Demographic composition of our Weibo user pool with 2018 Annual Sina Weibo user report

*COVID-19 Posts*

Following the best practices in content retrieval and analysis [27], we generated a comprehensive list of keywords related to COVID-19 through a close observation of Weibo posts daily from late January to early March 2020. We then retrieved COVID-19 posts by searching through all posts by users in the user pool, with these 167 keywords covering general terms related to epidemic, such as coronavirus and pneumonia, as well as specific locations (e.g., "Wuhan"), drugs (e.g., "remdesivir"), preventive measures (e.g., "mask"), among others (for a complete keyword list, see Multimedia Appendix: Table A).

After removing duplicates (i.e., reposts of original posts), we retained 14,983,647 posts sent between November 1st, 2019 (i.e., 30 days before the first confirmed cases) and March 31st, 2020.

A fraction of these posts (3.10%; N = 464,111) were tagged with geographic information. We distinguished between posts sent within Hubei province (i.e., the epicenter; N = 169,340, 36.49%) and those from elsewhere in mainland China (N = 294,771, 63.51%).

*Sick Posts*

We conceptually define "sick posts" as posts that report any symptoms and/or diagnoses that are likely related to COVID-19, based on published research and news reports from medical social media site DXY.cn [28]. We collected a broad list of symptoms, including common symptoms such as cough and shortness of breath, and uncommon symptoms such as diarrhea. Sick posts can be further categorized into "ingroup sick posts," which we define as posts that disclose one's own or immediate family members' symptoms or diagnoses, and "outgroup sick posts" which report those from other people not in one's immediate family. The reason for the *a priori* categorization is that people tend to have first-hand and more accurate information about their own or immediate family members' medical conditions, while people have much less reliable information about those outside of their household, especially during a national lockdown. All other posts that do not fall into these categories are classified as "other COVID-19 posts." We list one example of "ingroup sick post" and one for "outgroup sick post" below (translated and edited for brevity):

> Ingroup sick post: "During the SARS epidemic in 2003, I got pneumonia with symptoms of fever and cough, was suspected of being infected with SARS, and ended up being hospitalized for more than a month. Now we got COVID-19 in 2020 and I started coughing again, which has lasted for more than a month. What a mess <Face Palm>" (posted at 10:23 p.m., January 29th, 2020)

> Outgroup sick post: "One man in another village drank too much. He said he felt sick and had cold symptoms. His brother measured his temperature which turned out to be 38 Celsius. His brother called 120 and sent him to hospital. The whole village was shocked and everyone was afraid to go outside. "(posted at 10:14 p.m., January 29th, 2020)

We used supervised machine learning algorithms to identify sick posts from the keyword-retrieved COVID-19 posts. We first sampled 11,575 posts in proportion to the retrieved posts across 5 months of data collection. Next, 11 human judges annotated whether a post was an "ingroup sick post," "outgroup sick post," or "other COVID-19 post." The judges independently annotated a subset of 138 posts and achieved high agreement (Krippendorff's α = 0.945) before they divided and annotated the remaining posts. Then, the annotated posts were used to train machine learning models with various algorithms. Based on the classification performance (see

Table 1), we selected the model using the random forest algorithm (F1 score = .880). The model classified all COVID-19 posts into 394,658 (2.63%) "ingroup sick posts," 97,635 (0.65%) "outgroup sick posts," and 14,491,354 (96.71%) "other COVID-19 posts." Because of the scarcity of outgroup sick posts, we combined ingroup and outgroup sick posts in subsequent analyses.

**Table 1: Performance of Machine Learning Models Classifying Sick Posts**

|  | F1-measure | Precision | Accuracy | Recall |
|---|---:|---:|---:|---:|
| Decision Tree | 0.835 | 0.840 | 0.830 | 0.830 |
| Extra Tree | 0.785 | 0.785 | 0.785 | 0.785 |
| Extra Trees | 0.878 | 0.881 | 0.885 | 0.885 |
| K-nearest Neighbors | 0.810 | 0.819 | 0.819 | 0.819 |
| Multi-layer Perceptron | 0.847 | 0.845 | 0.851 | 0.851 |
| Support Vector Machine | 0.877 | 0.877 | 0.878 | 0.878 |
| Random Forest | 0.880 | 0.885 | 0.888 | 0.888 |

Among the subset of geotagged COVID-19 posts (3.10% of all retrieved posts), 5,650 sick posts and 163,690 other COVID-19 posts were tagged in Hubei, and 26,488 sick posts and 268,283 other COVID-19 posts were from elsewhere in mainland China. These post counts were then aggregated by days. To control for the day-to-day fluctuations of Weibo posts, we further normalized these numbers against the daily counts of all Weibo posts generated by our user pool. The normalized sick post and other COVID-19 post counts can be interpreted as counts per 1-million posts. Figure 2 summarizes our data collection and classification process.

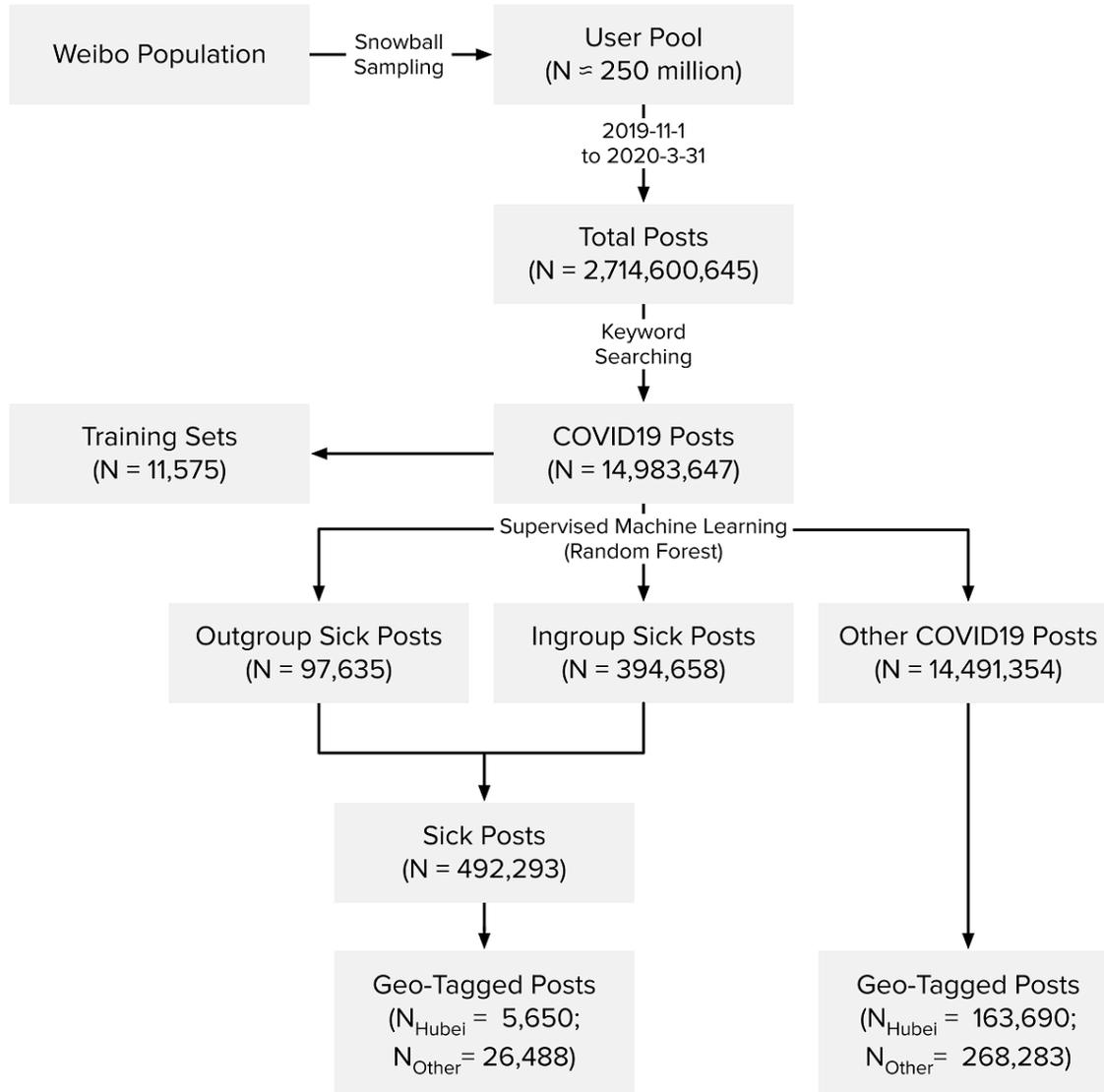

*Figure 2*. Weibo data collection and classification procedure

*COVID-19 Daily Case Counts*

We collected daily new case counts in mainland China from China CDC on May 8th, 2020. China CDC's official website started collating data on January 16th, 2020. Earlier counts were obtained from Huang et al. [1] and validated against relevant briefings from the National Health Commission. The final case data cover the same period from November 1st, 2019, to March 31st, 2020, within which the first reported COVID-19 clinical case dates back to December 1st, 2019. We also distinguished between the cases within and outside of Hubei (see Figure 3).

It is noteworthy that China CDC released seven editions of diagnostic criteria throughout the course covered in this study and thus introduced systematic changes to the case counts.

Particularly, on February 12[th], 2020, Hubei province started to implement the fifth edition released on February 4[th]. This led to a temporary surge of new cases [29]. The incident's impact was controlled for in our analyses, as discussed in the section below. After close inspection, we concluded that the changes among other editions of the diagnostic criteria were relatively minor and their release dates did not appear to be associated with abrupt changes of the case counts; therefore, we did not further control for them.

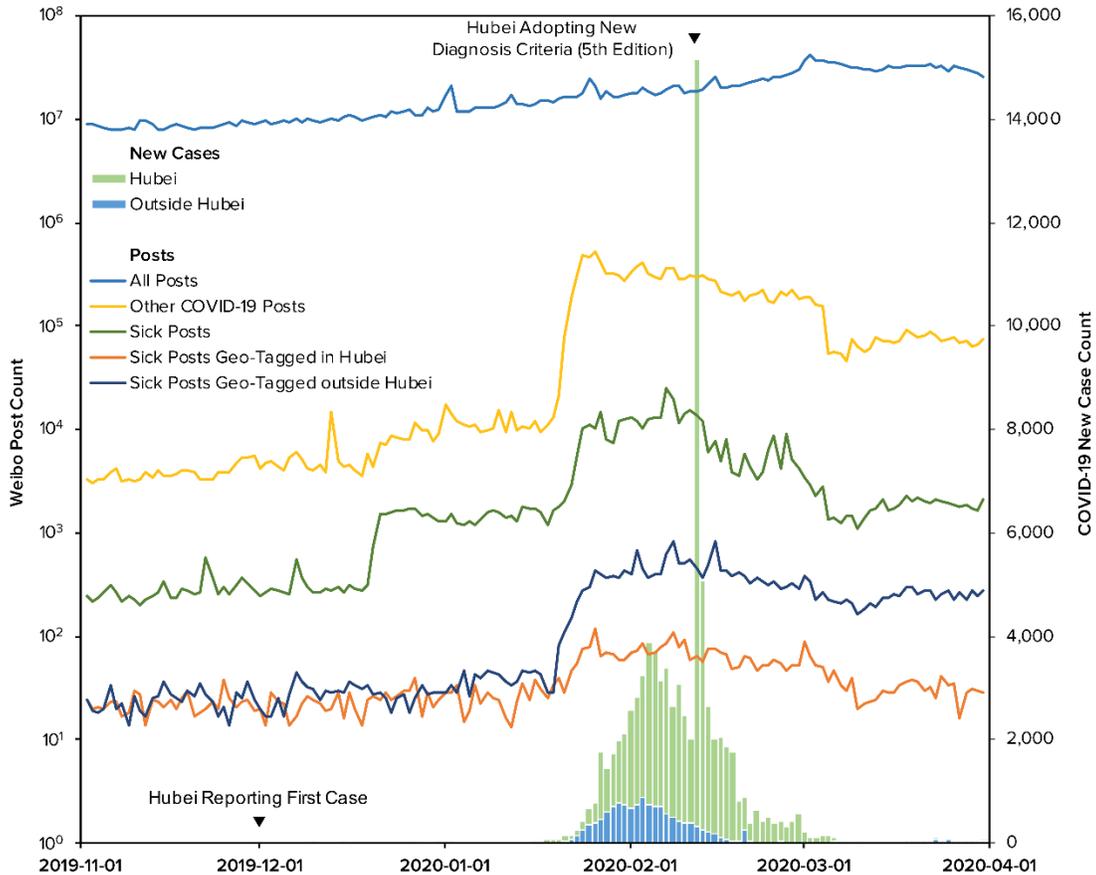

*Figure 3*. Daily Weibo posts and confirmed COVID-19 cases between November 1[st], 2019 and March 31[st], 2020

**Statistical Analysis**

We performed Granger causality tests [30] to discover if the increase of sick posts forecasts the increase of new cases, as formulated in the following linear model:

$$\Delta C_t = a_0 + \sum_{i=1}^{m} a_i \ \Delta C_{t-i} + \sum_{j=1}^{m} b_j \ \Delta S_{t-i} + c_1 I_t + \varepsilon_t$$

where $C_t$ is the difference of new case counts at day *t* from day *t-1*, $S_{t-i}$ is the difference of sick post counts (normalized) at day *t* from day *t-1*, and $I_t$ is a time-varying binary variable that equals 1 on February 12[th], 2020, on which day Hubei adopted the 5[th] edition of diagnostic criteria. This binary variable controls for the exogenous pulse of case counts [31]. Since Weibo posts were collected from as early as November 1st, 2019, 30 days before the first reported case on December 1st, 2019, we were able to test up to 29 lags of such posts (i.e., m ≤ 29). The model is further explained as follows.

First, difference scores instead of raw new case counts are used because Dickey-Fuller tests for the raw counts could not reject non-stationarity (i.e., the presence of a unit root) for lag 3–29 at 5% CL (confidence level; see Multimedia Appendix: Table B). Both stationarity and the inclusion of autoregressive terms are required by Granger causality. In contrast, the Dicky-Fuller tests suggest the difference scores of case counts are stationary: The non-stationarity is rejected for lag 1–12 at 1% CL, and it is rejected for lag 13–29 at 5% CL (see Multimedia Appendix: Table B). The Dickey-Fuller tests also reached the same conclusion for the stationarities of the sick post counts and their difference scores (see Multimedia Appendix: Table B). We thus also used the difference scores instead of the raw counts to reduce correlations among lag terms of sick post counts. This helps to better identify their independent effects on case counts. In short, these difference scores can be interpreted as "daily-additional" cases or Weibo posts beyond the counts from the previous day.

Second, to determine the number of lag terms to include (i.e., m), we compared model fit statistics while iteratively adding lag terms. The model comparison suggests the inclusion of more lags continuously improves model fit till the maximum lags (i.e., 29; see Multimedia Appendix: Table C). However, the parameter estimates do not change qualitatively after including more than 20 lags (see Multimedia Appendix: Tables D & E). For parsimony and statistical power, we settled at 20 lags for the following analyses.

Finally, we included a binary variable to control for the change of diagnosis criteria on Feb 12th, 2020, following the procedure of intervention analysis [32]. Because this change is unlikely to induce permanent change to case counts, an instant pulse function is applied at the date of change. We also tested models that allow the effect to linearly decay in 2, 3, 4, or 5 days, and these models fit the data worse than the model with an instant pulse (see Multimedia Appendix: Table F).

**Results**

Ordinary least square regressions with robust standard errors were used to estimate the final models. With 20 lag terms in the model, the dataset includes daily-additional new COVID-19 cases from December 1st, 2019 to March 31, 2020, and daily-additional counts of sick posts and other COVID-19 posts from November 10th, 2019 to March 11, 2020 (N = 122).

Figure 4(A) summarizes the estimates of Granger causality for sick posts predicting new COVID-19 cases with standardized regression coefficients (see Multimedia Appendix: Table G for all estimated parameters). Particularly, one standard deviation of increase in the daily-additional sick posts (1 sick post per 1-million posts) predicted 0.133 (95% CI: 0.065, 0.201) to 0.275 (95% CI: 0.134, 0.416) standard deviation of increase in the daily-additional new cases, 1 to 14 days in advance. After including the 20 lags of sick posts, the model's adjusted $R^2$ increases by 0.128, suggesting that sick posts could explain an additional 12.8% of the variance of daily-additional new cases beyond the autoregressive terms and intervention effects.

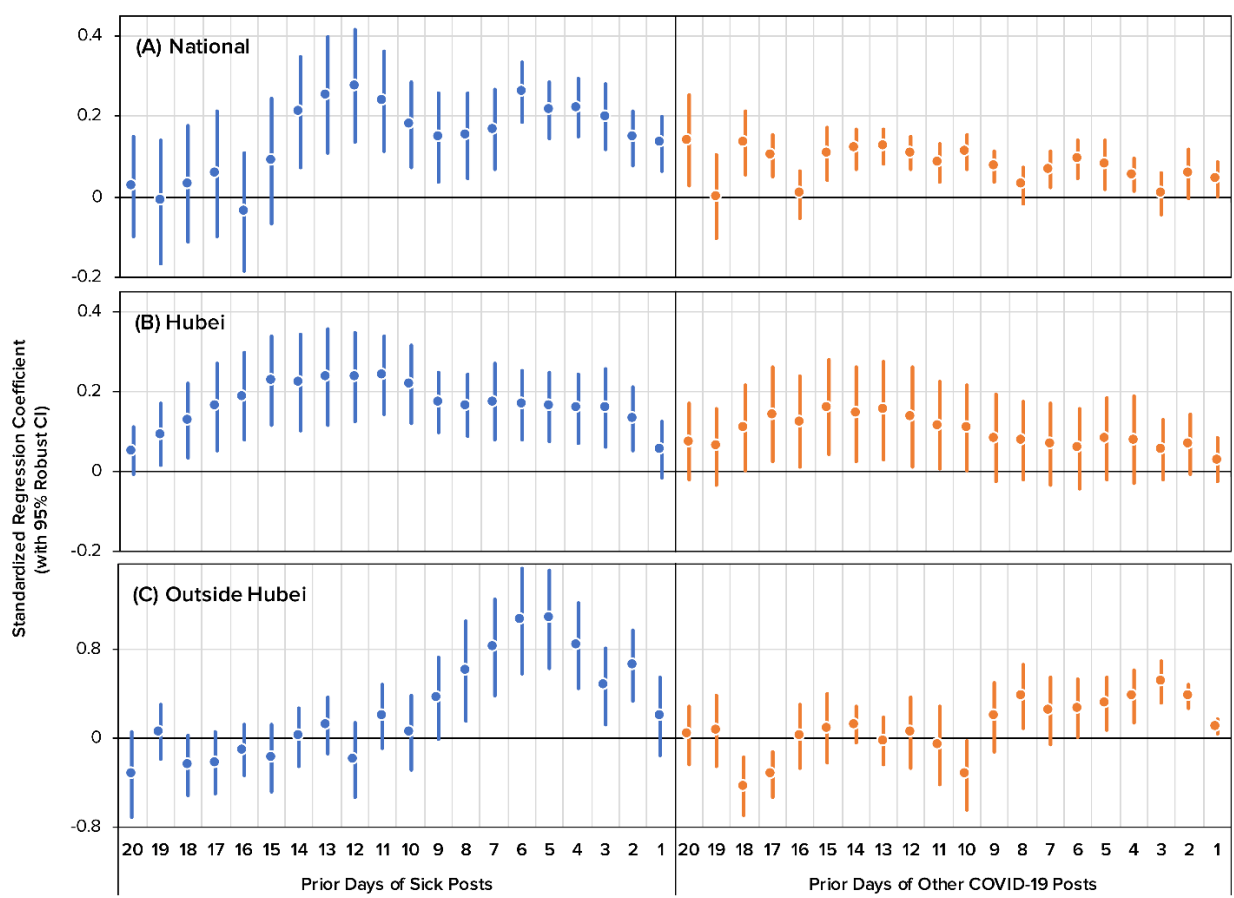

*Figure 4*. Standardized estimates of Granger causality for time-lagged, daily-additional Weibo posts (sick posts and other COVID-19 posts) predicting daily-additional cases.

Furthermore, we tested the relationship between daily-additional new cases and other COVID-19 post counts, using the same linear model. Figure 4(A) further illustrates the standardized coefficient estimates. Compared with sick posts, other COVID-19 posts were weaker signals of future case counts, as demonstrated by smaller standard regression coefficients. It means that Weibo posts that discuss some aspect of COVID-19 but did not report explicitly anyone's symptoms or diagnosis had a smaller forecasting power, compared with that of sick posts.

12To corroborate the above results, we tested sick posts' Granger causality on cases within Hubei and outside of Hubei (see Multimedia Appendix: Table H). Within Hubei, the results generally agree with the national pattern mentioned above. Daily-additional sick posts predicted daily-additional new cases within Hubei up to 19 days in advance, as illustrated in Figure 4(B). In contrast, other COVID-19 posts had fewer lag terms that could forecast new cases. Outside of Hubei, sick posts' predictive pattern was similar to the national pattern, but with a limited time range: sick posts could forecast new cases 2 to 8 days in advance (see Figure 4(C)).

**Discussion**

The novel coronavirus causing COVID-19 is a pathogen new to the human reservoir. It poses an extraordinary challenge for public health systems worldwide, because screening and diagnostic tests have to be developed from scratch. Even when such tests eventually become available, testing capacity is often severely limited, fueling the outbreak as many patients unknowingly infect others. Based on approximately 15 million COVID-19-related Weibo posts between November 1$^{st}$, 2019 and March 31$^{st}$, 2020, we developed a supervised machine learning classifier to identify "sick posts," which are reports of one's own and other people's symptoms and diagnosis of COVID-19. Using the officially reported daily case counts as the outcome, our work shows that sick posts significantly predict daily cases, up to 14 days ahead of official statistics. This finding confirms prior research that social media data can be usefully applied to nowcasting and forecasting emerging infectious diseases such as COVID-19 [22, 33].

One of the biggest challenges of digital disease surveillance is to identify useful disease signals, especially when facing a deluge of social media activities as a result of COVID-19 mitigation measures [12, 33-35]. Our finding that sick posts have a greater predictive power than other COVID-19 posts shows that not all social media data are equally informative. Specifically, COVID-19 has dramatically disrupted everyday life, resulting in people sheltering in place and increasingly communicating with others via social media. As shown in prior work [18] as well as in our dataset, the majority of COVID-19 chatter on Weibo was due to public awareness of COVID-19, rather than actual symptom reports. Most previous work took rather coarse-grained approaches, relying primarily on either aggregated search query data or social media data retrieved from limited keyword searches [19, 22]. Our work collects the largest, most comprehensive and granular social media data related to COVID-19 in the Chinese language. More importantly, it demonstrates one viable way to identify valid signals from noise through reports of symptoms and diagnosis, bringing significant contributions to the literature on digital surveillance.

Another important finding is that the predictive power of sick posts on daily case counts holds true for both Hubei and non-Hubei regions, but the effect sizes vary. Being the epicenter of the outbreak, Hubei province experienced extreme testing shortages during the early stage of the study period. As a result, many Hubei residents turned to social media sites such as Weibo to seek help for testing and medical care. By contrast, social media help-seeking activities were uncommon in other parts of China where testing and healthcare resources were much more adequate. With such regional variations, we still observed predictive signals of sick posts on case counts, suggesting that predictive power of sick posts was robust against testing delays. Further, the variations in the effect estimates show that social media data's predictive power may vary across different geographic areas, with different levels of preparedness, and at different stages of the outbreak. Future studies based on longer periods of data monitoring could explore in more depth the temporal and spatial variations of COVID-19 social media surveillance efficacy.

Our work has broad public health implications. The high speed and low cost of social media surveillance can be especially useful at the early stages of the COVID-19 outbreak, to inform containment and mitigation efforts when they are most cost-effective. For countries and regions where public health infrastructures do not allow for widespread screening and diagnostic tests, social media disease surveillance provides much needed information for public health agencies to model the trajectories of the outbreak, and make swift decisions about resource allocation, such as hospital beds, ventilators, and personal protective equipment.

Another advantage of social media surveillance is that it can be done from afar. As COVID-19 continues to spread across the globe, countries lacking testing and screening infrastructures will become "dark spots," endangering their own people as well as the entire world. It is imperative that international organizations such as the World Health Organization integrate such data into their outbreak forecasting management practices, in order to mobilize and coordinate relief efforts to help combat COVID-19.

This study has several limitations. First, Weibo posts were retrieved retrospectively, rather than in real-time, which means deleted or censored posts were absent from our dataset. However, we have no reason to believe that deletion or censorship favored "sick posts" in measurable ways. In fact, a recent study on Weibo censorship during December 2019 to February 2020 shows that only 1.7 per 1000 Weibo posts were censored, and these censored posts were generally about the government's missteps in its COVID-19 response, not individual reports of symptoms and diagnoses [36]. Therefore, our results should be unaffected by censorship omissions. Second, as some studies suggest [37-39], confirmed COVID-19 case counts published by China CDC may be a significant underestimation of the actual counts, due in part to limits in testing capacity and the existence of asymptomatic carriers. Still, the data here represents the best-known data of confirmed case counts. Third, it is important to acknowledge that "sick posts," as disease signals, are not without noises, because 1) Weibo users who reported COVID-19 symptoms did not necessarily have COVID-19 clinically, 2) Weibo users may not speak the truth, and 3) Weibo users may "overreport" (posting about their symptoms or diagnoses multiple times) or "underreport" (not posting despite their symptoms or diagnoses) for a variety of reasons. Such inaccuracies are inherent in user-generated social media data, and widely exist in all infoveillance studies. However, it should be noted that the goal of infoveillance has never been to achieve one-for-one match between social media posts and clinical cases. Rather, infoveillance approaches strive to mine useful, early signals from social media and Internet data as a supplement to conventional surveillance efforts. Despite such noises, we still found these sick posts reliably predicted COVID-19 case counts, indicating the signal's validity in reflecting disease spread in the population.

The threats of COVID-19 and other infectious diseases are likely to recur in the future. Reports of symptoms and diagnosis on social media during emerging disease outbreaks send invaluable warning signals to the public. Researchers and disease control agencies should pay close attention to the social media infosphere. On top of monitoring overall search and posting activities, it is crucial to sift through the contents and efficiently identify true signals from noise. Our main findings highlight the importance of using rigorous procedures and understanding information sharing behaviors to obtain quality signals to quantify sickness reports. Future studies based on longer periods of data monitoring could explore in more depth the time and spatial diffusion of COVID-19. More detailed examination of post contents reporting restraints in information or medical resources will be helpful in developing local outbreak responses.

**Acknowledgements**

CS, WL, JZ, and BF contributed to the study design. AC collected Weibo data. WL, CL and AC contributed to data analysis. WL, CS, CL and AC contributed to the design and drawing of figures. All authors contributed to the writing of the manuscript. We thank Jingyang Xu, Minwei Ren, Rixia Tang, Zichao Wang, Yongyan Xu, Na yang, Yalan Jin, Xiuchan Xu, Xinyu Wang, Ruizhi Sun, Wenhui Zhu, Yiwei Li, Tianyu Zhao for their help with data annotation.

**Conflicts of Interest**

We declare no conflicts of interest.

**Multimedia Appendix**

Table A. COVID-19 related keywords used to retrieve Weibo posts
Table B. Summaries of modified Dickey-Fuller t tests for a unit root (without trend) in the time series of new cases, sick posts, or other COVID-19 posts in mainland China (N = 122)
Table C. Model comparisons for sick post or other COVID-19 post (difference scores) predicting new cases (difference scores) in mainland China with varying lag terms (N = 122)
Table D. Model summaries for sick posts (difference scores) predicting new cases (difference scores) in mainland China with varying lag terms (N = 122)
Table E. Model summaries for other COVID-19 posts (difference scores) predicting new cases (difference scores) in mainland China with varying lag terms (N = 122)

Table F. Model comparisons for sick posts or other COVID-19 posts (difference scores) predicting new cases (difference scores) in mainland China with varying linear decay rates of the effect of the changed diagnostic criteria on February 12th, 2020 (N = 122)
Table G. Model summaries of sick posts or other COVID-19 posts (difference scores) predicting new cases (difference scores) in mainland China, including a baseline model without effects of social media posts (N = 122)
Table H. Model summaries of sick posts or other COVID-19 posts (difference scores) predicting new cases (difference scores) within or outside Hubei (N = 122)

**Using Reports of Own and Others' Symptoms and Diagnosis on Social Media to Predict COVID-19 Case Counts: Observational Infoveillance Study in Mainland China**

**Multimedia Appendix**

**Table A. COVID-19 related keywords used to retrieve Weibo posts**

| Keyword | Translation | Keyword | Translation |
| --- | --- | --- | --- |
| 武汉肺炎 | Wuhan pneumonia | 潜伏期 | Incubation period |
| 新型冠状病毒肺炎 | COVID-19 | 北京 AND 病例 | Beijing AND Cases |
| 不明原因肺炎 | Pneumonia of unknown cause | 天津 AND 病例 | Tianjin AND Cases |
| 肺炎疫情 | Pneumonia outbreak | 河北 AND 病例 | Hebei AND Cases |
| 野味肺炎 | Wildlife pneumonia | 辽宁 AND 病例 | Liaoning AND Cases |
| 新型冠状病毒 AND 确诊 | Novel coronavirus AND Confirmed infected | 上海 AND 病例 | Shanghai AND Cases |
| 感染人数 | Number of infected cases | 江苏 AND 病例 | Jiangsu AND Cases |
| 出门 AND 戴口罩 | Going out AND Wear mask | 浙江 AND 病例 | Zhejiang AND Cases |
| N95 AND 口罩 | N95 AND Mask | 福建 AND 病例 | Fujian AND Cases |
| 3M AND 口罩 | 3M AND Mask | 山东 AND 病例 | Shandong AND Cases |
| KN95 AND 口罩 | KN95 AND Mask | 广东 AND 病例 | Guangdong AND Cases |
| 大众畜牧野味店 | Dazhong wildlife restaurant | 海南 AND 病例 | Hainan AND Cases |
| 口罩 | Mask | 山西 AND 病例 | Shanxi AND Cases |
| 新肺炎 | Novel pneumonia | 内蒙古 AND 病例 | Inner Mongolia AND Cases |
| 华南野生市场 | South China wild market | 吉林 AND 病例 | Jilin AND Cases |
| 冠状肺炎 | Corona pneumonia | 黑龙江 AND 病例 | Heilongjiang AND Cases |
| 武汉病毒所 | Wuhan Institute of Virology | 安徽 AND 病例 | Anhui AND Cases |
| China AND CDC | China AND Center for Disease Control and Prevention | 江西 AND 病例 | Jiangxi AND Cases |
| 中国疾病预防控制中心 | Chinese Center for Disease Control and Prevention | 河南 AND 病例 | Henan AND Cases |
| #2019nCoV | .. | 湖北 AND 病例 | Hubei AND Cases |
| 双黄连 AND 抢购 | Shuanghuanglian AND Rush to buy | 湖南 AND 病例 | Hunan AND Cases |
| 双黄连 AND 售罄 | Shuanghuanglian AND Sold out | 广西 AND 病例 | Guangxi AND Cases |
| 武汉卫健委 | Wuhan Municipal Health Committee | 四川 AND 病例 | Sichuan AND Cases |
| 湖北卫健委 | Health Commission of Hubei Province | 贵州 AND 病例 | Guizhou AND Cases |
| 肺炎 | Pneumonia | 云南 AND 病例 | Yunnan AND Cases |
| 疫情 | Epidemic outbreak | 西藏 AND 病例 | Tibet AND Cases |
| 隔离 | Quarantine | 陕西 AND 病例 | Shanxi AND Cases |
| 火神山 | Huoshen Shan hospital | 甘肃 AND 病例 | Gansu AND Cases |
| 雷神山 | Leishen Shan hospital | 青海 AND 病例 | Qinghai AND Cases |
| 钟南山 | Zhong Nanshan | 宁夏 AND 病例 | Ningxia AND Cases |
| 疫情防控 | Epidemic prevention and control | 新疆 AND 病例 | Xinjiang AND Cases |
| Coronavirus | .. | 香港 AND 病例 | Hong Kong AND Cases |
| Remdesivir | .. | 澳门 AND 病例 | Macau AND Cases |
| 瑞德西韦 | Remdesivir | 台湾 AND 病例 | Taiwan AND Cases |
| 新型肺炎 AND 死亡 | Novel coronavirus pneumonia AND Death | ECMO | Extracorporeal Membrane Oxygenation |
| 新型肺炎 AND 感染 | Novel coronavirus pneumonia AND Infection | 人工膜肺 | Extracorporeal membrane oxygenation |
| 新型冠状病毒 AND 感染 | Novel coronavirus AND Infection | 双盲测试 | Double blind test |
| 感染 AND 案例 | Infected AND Cases | 核酸检测 | Nucleic acid testing |
| 武汉 AND 封城 | Wuhan AND Lockdown | 疫苗 | Vaccine |
| 高福 | George Fu Gao | 小区出入证 | Community pass card |
| 王延轶 | Wang Yanyi | 战疫 | Anti-COVID-19 |
| 舒红兵 | Shu Hongbing | 抗疫 | Anti-COVID-19 |
| 协和医院 | Xiehe Hospital | 全国疫情 | Epidemic in China |
| 武汉 AND 隔离 | Wuhan AND Quarantine | 囤积口罩 | Hoarding mask |
| 医生 AND 李文亮 | Doctor AND Li Wenliang | 湖北卫健委 AND 免职 | Health commission of Hubei Province AND Remove from the position |
| 云监工 | Supervising work on cloud | 发热患者 | Fever patients |
| 武汉 AND 肺炎 AND 谣言 | Wuhan AND Pneumonia AND Rumors | 延迟开学 | Postpone the reopening of school |
| 8 名 AND 散布武汉肺炎谣言 | Eight people AND Spread rumors of Wuhan pneumonia | 开学时间 AND 不得早于 | The start time of school AND Not earlier than |

| Keyword | Translation | Keyword | Translation |
|---|---|---|---|
| 武汉仁爱医院 | Wuhan Ren'ai Hospital | 累计死亡数 | Cumulative deaths |
| 黄冈 AND 新肺炎 | Huanggang AND Novel pneumonia | 疑似病例 | Suspicious cases |
| 黄冈 AND 新型冠状病毒 | Huanggang AND Novel coronavirus | 入户排查 | Household troubleshoot |
| 黄冈 AND 感染者 | Huanggang AND Infected cases | 武汉市慈善总会 | Wuhan Charity Federation |
| 孝感 AND 新肺炎 | Xiaogan AND Novel pneumonia | 防疫物资 | Epidemic control and prevention materials |
| 孝感 AND 新型冠状病毒 | Xiaogan AND Novel coronavirus | 捐赠物资 | Donation materials |
| 孝感 AND 感染者 | Xiaogan AND Infected cases | 俄罗斯 AND 捐赠 | Russia AND Donations |
| 居家隔离 | Isolated at home | 巴基斯坦 AND 捐赠 | Pakistan AND Donations |
| 隔离 AND 14 天 | Isolation AND 14 days | 美国 AND 捐赠 | United States AND Donations |
| 潜伏期 AND 24 天 | Incubation period AND 24 days | 日本 AND 捐赠 | Japan AND Donations |
| 潜伏期 AND 14 天 | Incubation period AND 14 days | MERS | Middle East Respiratory Syndrome |
| 新型肺炎 | Novel pneumonia | 中央赴湖北指导小组 | Delegation from central government to guide Hubei |
| 新型冠状病毒 | Novel coronavirus | 抗击 AND 新型肺炎 | Fight against AND COVID-19 |
| 国际公共卫生紧急事件 | International Public Health Emergencies | 支援武汉 | Give a hand to Wuhan |
| PHEIC | International Public Health Emergencies | 医用口罩 | Surgical mask |
| #nCoV | .. | 武汉 AND 新增 | Wuhan AND Novel cases |
| 方舱医院 | FangCang Hospital | 临床诊断病例 | Clinically diagnosed cases |
| 一省包一市 | One province gives a hand to one Hubei city | 应勇 AND 湖北 | Ying Yong AND Hubei |
| 新冠肺炎 | Novel coronavirus pneumonia | 应勇 AND 上海 | Ying Yong AND Shanghai |
| 晋江毒王 | Super spreader of COVID-19 in Jinjiang | 蒋超良 AND 湖北 | Jiang Chaoliang AND Hubei |
| 超级传播者 | Super spreader | SARS-CoV-2 | .. |
| 湖北 AND 王晓东 | Hubei AND Wang Xiaodong | 武汉 AND 死亡病例 | Wuhan AND Death cases |
| 蒋超良 | Jiang Chaoliang | 武汉 AND 感染病例 | Wuhan AND Infection cases |
| #武汉肺炎 | #Wuhan pneumonia | 湖北 AND 死亡病例 | Hubei AND Death cases |
| 武汉 AND 李文亮 | Wuhan AND Li Wenliang | 湖北 AND 感染病例 | Hubei AND Infected cases |
| 武汉 AND 李医生 | Wuhan AND Dr. Li | 中国 AND 死亡病例 | China AND Death cases |
| 武汉 AND 疫情 | Wuhan AND Epidemic | 中国 AND 感染病例 | China AND Infected cases |
| 国家疾控中心 | Chinese Center for Disease Control and Prevention | 企业复工 | Enterprise work resuming |
| 武汉 AND 疫苗 | Wuhan AND Vaccine | 中小企业 AND 困境 | Small and medium-sized enterprise AND Dilemma |
| 管轶 | Guan Yi | 超市采购 | Supermarket Purchase |
| 张晋 AND 卫健委 | Zhang Jin AND Health Commission | 西贝 | Xibei |
| 张晋 AND 卫生健康委员会 | Zhang Jin AND Health Commission | 武汉 AND 征用宿舍 | Wuhan AND Requisitioned students' dormitory |
| 刘英姿 AND 卫健委 | Liu Yingzi AND Health Commission | 周佩仪 | Zhou Peiyi |
| 刘英姿 AND 卫生健康委员会 | Liu Yingzi AND Health Commission | 武汉中心医院 | The Central Hospital of Wuhan |
| 王贺胜 AND 卫健委 | Wang Hesheng AND Health Commission | 武汉病毒研究 | Virology research in Wuhan |
| 王贺胜 AND 卫生健康委员会 | Wang Hesheng AND Health Commission | | |

**Table B. Summaries of modified Dickey-Fuller *t* tests for a unit root (without trend) in the time series of new cases, sick posts, or other COVID-19 posts in mainland China ($N = 122$)**

| Max Lags | New Cases Raw Counts | | | | New Cases Difference Scores (Daily-Additional Counts) | | | | Sick Posts Raw Counts | | | | Sick Posts Difference Scores (Daily-Additional Counts) | | | | Other COVID-19 Posts Raw Counts | | | | Other COVID-19 Posts Difference Scores (Daily-Additional Counts) | | | |
|---|---|---|---|---|---|---|---|---|---|---|---|---|---|---|---|---|---|---|---|---|---|---|---|---|
| | *t* | 1% CV | 5% CV | 10% CV | *t* | 1% CV | 5% CV | 10% CV | *t* | 1% CV | 5% CV | 10% CV | *t* | 1% CV | 5% CV | 10% CV | *t* | 1% CV | 5% CV | 10% CV | *t* | 1% CV | 5% CV | 10% CV |
| 29 | -1.212 | -2.597 | -1.899 | -1.557 | -2.087 | -2.597 | -1.899 | -1.557 | -1.024 | -2.597 | -1.899 | -1.557 | -1.993 | -2.597 | -1.899 | -1.557 | -1.075 | -2.597 | -1.899 | -1.557 | -1.833 | -2.597 | -1.899 | -1.557 |
| 28 | -1.234 | -2.597 | -1.887 | -1.551 | -2.139 | -2.597 | -1.887 | -1.551 | -0.989 | -2.597 | -1.887 | -1.551 | -1.940 | -2.597 | -1.887 | -1.551 | -1.207 | -2.597 | -1.887 | -1.551 | -1.829 | -2.597 | -1.887 | -1.551 |
| 27 | -1.264 | -2.597 | -1.879 | -1.548 | -2.186 | -2.597 | -1.879 | -1.548 | -0.962 | -2.597 | -1.879 | -1.548 | -2.091 | -2.597 | -1.879 | -1.548 | -1.134 | -2.597 | -1.879 | -1.548 | -1.677 | -2.597 | -1.879 | -1.548 |
| 26 | -1.304 | -2.597 | -1.873 | -1.546 | -2.222 | -2.597 | -1.873 | -1.546 | -1.024 | -2.597 | -1.873 | -1.546 | -2.250 | -2.597 | -1.873 | -1.546 | -1.110 | -2.597 | -1.873 | -1.546 | -1.839 | -2.597 | -1.873 | -1.546 |
| 25 | -1.349 | -2.597 | -1.869 | -1.547 | -2.242 | -2.597 | -1.869 | -1.547 | -1.070 | -2.597 | -1.869 | -1.547 | -2.209 | -2.597 | -1.869 | -1.547 | -1.206 | -2.597 | -1.869 | -1.547 | -1.942 | -2.597 | -1.869 | -1.547 |
| 24 | -1.402 | -2.597 | -1.868 | -1.549 | -2.252 | -2.597 | -1.868 | -1.549 | -1.110 | -2.597 | -1.868 | -1.549 | -2.206 | -2.597 | -1.868 | -1.549 | -1.124 | -2.597 | -1.868 | -1.549 | -1.842 | -2.597 | -1.868 | -1.549 |
| 23 | -1.464 | -2.597 | -1.868 | -1.553 | -2.249 | -2.597 | -1.868 | -1.553 | -1.091 | -2.597 | -1.868 | -1.553 | -2.216 | -2.597 | -1.868 | -1.553 | -1.195 | -2.597 | -1.868 | -1.553 | -2.047 | -2.597 | -1.868 | -1.553 |
| 22 | -1.521 | -2.597 | -1.871 | -1.558 | -2.230 | -2.597 | -1.871 | -1.558 | -1.251 | -2.597 | -1.871 | -1.558 | -2.357 | -2.597 | -1.871 | -1.558 | -1.267 | -2.597 | -1.871 | -1.558 | -1.990 | -2.597 | -1.871 | -1.558 |
| 21 | -1.578 | -2.597 | -1.875 | -1.565 | -2.219 | -2.597 | -1.875 | -1.565 | -1.371 | -2.597 | -1.875 | -1.565 | -2.135 | -2.597 | -1.875 | -1.565 | -1.411 | -2.597 | -1.875 | -1.565 | -1.934 | -2.597 | -1.875 | -1.565 |
| 20 | -1.641 | -2.597 | -1.881 | -1.574 | -2.206 | -2.597 | -1.881 | -1.574 | -1.290 | -2.597 | -1.881 | -1.574 | -2.009 | -2.597 | -1.881 | -1.574 | -1.444 | -2.597 | -1.881 | -1.574 | -1.780 | -2.597 | -1.881 | -1.574 |
| 19 | -1.673 | -2.597 | -1.888 | -1.583 | -2.184 | -2.597 | -1.888 | -1.583 | -1.428 | -2.597 | -1.888 | -1.583 | -2.206 | -2.597 | -1.888 | -1.583 | -1.238 | -2.597 | -1.888 | -1.583 | -1.777 | -2.597 | -1.888 | -1.583 |
| 18 | -1.697 | -2.597 | -1.897 | -1.594 | -2.200 | -2.597 | -1.897 | -1.594 | -1.578 | -2.597 | -1.897 | -1.594 | -2.056 | -2.597 | -1.897 | -1.594 | -1.355 | -2.597 | -1.897 | -1.594 | -2.139 | -2.597 | -1.897 | -1.594 |
| 17 | -1.693 | -2.597 | -1.907 | -1.605 | -2.229 | -2.597 | -1.907 | -1.605 | -1.712 | -2.597 | -1.907 | -1.605 | -1.905 | -2.597 | -1.907 | -1.605 | -1.388 | -2.597 | -1.907 | -1.605 | -2.012 | -2.597 | -1.907 | -1.605 |
| 16 | -1.663 | -2.597 | -1.918 | -1.617 | -2.298 | -2.597 | -1.918 | -1.617 | -1.629 | -2.597 | -1.918 | -1.617 | -1.787 | -2.597 | -1.918 | -1.617 | -1.548 | -2.597 | -1.918 | -1.617 | -2.019 | -2.597 | -1.918 | -1.617 |
| 15 | -1.630 | -2.597 | -1.931 | -1.630 | -2.414 | -2.597 | -1.931 | -1.630 | -1.600 | -2.597 | -1.931 | -1.630 | -1.911 | -2.597 | -1.931 | -1.630 | -1.287 | -2.597 | -1.931 | -1.630 | -1.851 | -2.597 | -1.931 | -1.630 |
| 14 | -1.691 | -2.597 | -1.943 | -1.644 | -2.552 | -2.597 | -1.943 | -1.644 | -1.466 | -2.597 | -1.943 | -1.644 | -1.986 | -2.597 | -1.943 | -1.644 | -1.150 | -2.597 | -1.943 | -1.644 | -2.299 | -2.597 | -1.943 | -1.644 |
| 13 | -1.660 | -2.597 | -1.957 | -1.658 | -2.551 | -2.597 | -1.957 | -1.658 | -1.284 | -2.597 | -1.957 | -1.658 | -2.224 | -2.597 | -1.957 | -1.658 | -1.096 | -2.597 | -1.957 | -1.658 | -2.701 | -2.597 | -1.957 | -1.658 |
| 12 | -1.628 | -2.597 | -1.971 | -1.672 | -2.701 | -2.597 | -1.971 | -1.672 | -1.304 | -2.597 | -1.971 | -1.672 | -2.643 | -2.597 | -1.971 | -1.672 | -1.497 | -2.597 | -1.971 | -1.672 | -3.004 | -2.597 | -1.971 | -1.672 |
| 11 | -1.583 | -2.597 | -1.986 | -1.686 | -2.880 | -2.597 | -1.986 | -1.686 | -1.180 | -2.597 | -1.986 | -1.686 | -2.725 | -2.597 | -1.986 | -1.686 | -1.495 | -2.597 | -1.986 | -1.686 | -2.276 | -2.597 | -1.986 | -1.686 |
| 10 | -1.526 | -2.597 | -2.000 | -1.701 | -3.122 | -2.597 | -2.000 | -1.701 | -1.206 | -2.597 | -2.000 | -1.701 | -3.199 | -2.597 | -2.000 | -1.701 | -1.397 | -2.597 | -2.000 | -1.701 | -2.353 | -2.597 | -2.000 | -1.701 |
| 9 | -1.475 | -2.597 | -2.015 | -1.715 | -3.456 | -2.597 | -2.015 | -1.715 | -1.449 | -2.597 | -2.015 | -1.715 | -3.349 | -2.597 | -2.015 | -1.715 | -1.276 | -2.597 | -2.015 | -1.715 | -2.619 | -2.597 | -2.015 | -1.715 |
| 8 | -1.455 | -2.597 | -2.030 | -1.729 | -3.881 | -2.597 | -2.030 | -1.729 | -1.686 | -2.597 | -2.030 | -1.729 | -2.937 | -2.597 | -2.030 | -1.729 | -1.126 | -2.597 | -2.030 | -1.729 | -3.022 | -2.597 | -2.030 | -1.729 |
| 7 | -1.416 | -2.597 | -2.044 | -1.743 | -4.351 | -2.597 | -2.044 | -1.743 | -1.477 | -2.597 | -2.044 | -1.743 | -2.623 | -2.597 | -2.044 | -1.743 | -1.244 | -2.597 | -2.044 | -1.743 | -3.707 | -2.597 | -2.044 | -1.743 |
| 6 | -1.515 | -2.597 | -2.059 | -1.757 | -5.104 | -2.597 | -2.059 | -1.757 | -1.423 | -2.597 | -2.059 | -1.757 | -3.149 | -2.597 | -2.059 | -1.757 | -1.211 | -2.597 | -2.059 | -1.757 | -3.618 | -2.597 | -2.059 | -1.757 |
| 5 | -1.592 | -2.597 | -2.073 | -1.770 | -5.563 | -2.597 | -2.073 | -1.770 | -1.220 | -2.597 | -2.073 | -1.770 | -3.489 | -2.597 | -2.073 | -1.770 | -1.338 | -2.597 | -2.073 | -1.770 | -4.071 | -2.597 | -2.073 | -1.770 |
| 4 | -1.729 | -2.597 | -2.086 | -1.782 | -6.413 | -2.597 | -2.086 | -1.782 | -1.088 | -2.597 | -2.086 | -1.782 | -4.526 | -2.597 | -2.086 | -1.782 | -1.345 | -2.597 | -2.086 | -1.782 | -4.025 | -2.597 | -2.086 | -1.782 |
| 3 | -2.049 | -2.597 | -2.098 | -1.793 | -7.564 | -2.597 | -2.098 | -1.793 | -1.214 | -2.597 | -2.098 | -1.793 | -6.065 | -2.597 | -2.098 | -1.793 | -1.574 | -2.597 | -2.098 | -1.793 | -4.433 | -2.597 | -2.098 | -1.793 |
| 2 | -2.443 | -2.597 | -2.110 | -1.804 | -8.575 | -2.597 | -2.110 | -1.804 | -1.146 | -2.597 | -2.110 | -1.804 | -6.686 | -2.597 | -2.110 | -1.804 | -1.179 | -2.597 | -2.110 | -1.804 | -4.142 | -2.597 | -2.110 | -1.804 |
| 1 | -3.221 | -2.597 | -2.120 | -1.814 | -10.731 | -2.597 | -2.120 | -1.814 | -1.642 | -2.597 | -2.120 | -1.814 | -10.739 | -2.597 | -2.120 | -1.814 | -1.187 | -2.597 | -2.120 | -1.814 | -6.801 | -2.597 | -2.120 | -1.814 |

Note. CV = critical value.

**Table C. Model comparisons for sick post or other COVID-19 post (difference scores) predicting new cases (difference scores) in mainland China with varying lag terms (*N* = 122)**

| | Cases Regressed on Sick Posts | | | | Cases Regressed on Other COVID-19 Posts | | | |
|---|---|---|---|---|---|---|---|---|
| **Max Lags** | **Adjusted R² (Δ)** | **AIC** | **BIC** | **Model df** | **Adjusted R² (Δ)** | **AIC** | **BIC** | **Model df** |
| 1 | 0.766 (–) | 1963.334 | 1974.550 | 3 | 0.746 (.) | 1973.108 | 1984.324 | 3 |
| 2 | 0.820 (0.055) | 1932.878 | 1949.702 | 5 | 0.804 (0.058) | 1943.581 | 1960.405 | 5 |
| 3 | 0.870 (0.050) | 1895.064 | 1917.497 | 7 | 0.831 (0.027) | 1927.413 | 1949.845 | 7 |
| 4 | 0.883 (0.013) | 1884.280 | 1912.320 | 9 | 0.843 (0.012) | 1920.511 | 1948.551 | 9 |
| 5 | 0.900 (0.017) | 1866.985 | 1900.633 | 11 | 0.846 (0.004) | 1919.511 | 1953.160 | 11 |
| 6 | 0.929 (0.029) | 1826.665 | 1865.921 | 13 | 0.850 (0.004) | 1918.410 | 1957.666 | 13 |
| 7 | 0.934 (0.005) | 1819.595 | 1864.460 | 15 | 0.853 (0.003) | 1917.858 | 1962.723 | 15 |
| 8 | 0.936 (0.002) | 1817.837 | 1868.309 | 17 | 0.852 (-0.001) | 1920.235 | 1970.708 | 17 |
| 9 | 0.936 (0.000) | 1820.310 | 1876.390 | 19 | 0.853 (0.001) | 1921.165 | 1977.245 | 19 |
| 10 | 0.936 (0.001) | 1820.552 | 1882.241 | 21 | 0.862 (0.010) | 1914.515 | 1976.203 | 21 |
| 11 | 0.939 (0.002) | 1817.247 | 1884.544 | 23 | 0.868 (0.006) | 1910.896 | 1978.192 | 23 |
| 12 | 0.946 (0.008) | 1802.692 | 1875.596 | 25 | 0.882 (0.014) | 1898.750 | 1971.655 | 25 |
| 13 | 0.949 (0.003) | 1797.454 | 1875.966 | 27 | 0.887 (0.005) | 1894.797 | 1973.310 | 27 |
| 14 | 0.957 (0.008) | 1777.530 | 1861.651 | 29 | 0.895 (0.008) | 1887.346 | 1971.466 | 29 |
| 15 | 0.964 (0.007) | 1757.691 | 1847.420 | 31 | 0.913 (0.019) | 1864.928 | 1954.657 | 31 |
| 16 | 0.964 (0.000) | 1758.488 | 1853.824 | 33 | 0.917 (0.004) | 1861.019 | 1956.355 | 33 |
| 17 | 0.966 (0.002) | 1751.786 | 1852.731 | 35 | 0.930 (0.013) | 1842.193 | 1943.138 | 35 |
| 18 | 0.969 (0.002) | 1743.840 | 1850.392 | 37 | 0.942 (0.012) | 1819.755 | 1926.308 | 37 |
| 19 | 0.970 (0.001) | 1742.406 | 1854.566 | 39 | 0.947 (0.005) | 1809.707 | 1921.868 | 39 |
| 20 | 0.970 (0.000) | 1741.721 | 1859.490 | 41 | 0.954 (0.007) | 1792.342 | 1910.110 | 41 |
| 21 | 0.973 (0.003) | 1731.055 | 1854.432 | 43 | 0.960 (0.006) | 1776.301 | 1899.678 | 43 |
| 22 | 0.973 (0.000) | 1729.606 | 1858.591 | 45 | 0.964 (0.004) | 1765.199 | 1894.184 | 45 |
| 23 | 0.973 (0.000) | 1731.977 | 1866.570 | 47 | 0.972 (0.008) | 1737.516 | 1872.109 | 47 |
| 24 | 0.974 (0.001) | 1726.073 | 1866.274 | 49 | 0.980 (0.009) | 1694.121 | 1834.322 | 49 |
| 25 | 0.978 (0.004) | 1706.914 | 1852.723 | 51 | 0.984 (0.004) | 1665.321 | 1811.130 | 51 |
| 26 | 0.982 (0.004) | 1683.031 | 1834.448 | 53 | 0.986 (0.001) | 1655.912 | 1807.329 | 53 |
| 27 | 0.983 (0.001) | 1676.765 | 1833.790 | 55 | 0.988 (0.002) | 1637.829 | 1794.854 | 55 |
| 28 | 0.984 (0.001) | 1670.747 | 1833.380 | 57 | 0.989 (0.001) | 1627.644 | 1790.277 | 57 |
| 29 | 0.986 (0.002) | 1650.172 | 1818.414 | 59 | 0.988 (0.000) | 1630.254 | 1798.496 | 59 |

**Table D. Model summaries for sick posts (difference scores) predicting new cases (difference scores) in mainland China with varying lag terms ($N = 122$)**

| | Model 1 (max lag = 5) | | | Model 2 (max lag = 10) | | | Model 3 (max lag = 15) | | | Model 4 (max lag = 20) | | | Model 5 (max lag = 25) | | | Model 6 (max lag = 29) | | |
|---|---|---|---|---|---|---|---|---|---|---|---|---|---|---|---|---|---|---|
| | *B* | *SE* | *p* | *B* | *SE* | *p* | *B* | *SE* | *p* | *B* | *SE* | *p* | *B* | *SE* | *p* | *B* | *SE* | *p* |
| Intercept | -114.238 | 44.247 | 0.011 | -99.007 | 35.435 | 0.006 | -102.012 | 26.627 | 0.000 | -99.941 | 24.650 | 0.000 | -91.168 | 21.238 | 0.000 | -93.310 | 16.901 | 0.000 |
| Change of Diagnosis Criteria | 13703.510 | 565.800 | 0.000 | 11712.520 | 553.424 | 0.000 | 11649.240 | 447.200 | 0.000 | 11476.130 | 612.788 | 0.000 | 10474.620 | 630.933 | 0.000 | 10594.030 | 550.274 | 0.000 |
| **Daily Additional New Cases** | | | | | | | | | | | | | | | | | | |
| Lag = 1 | -0.620 | 0.036 | 0.000 | -0.682 | 0.045 | 0.000 | -0.746 | 0.037 | 0.000 | -0.686 | 0.049 | 0.000 | -0.695 | 0.051 | 0.000 | -0.739 | 0.045 | 0.000 |
| Lag = 2 | -0.490 | 0.043 | 0.000 | -0.552 | 0.059 | 0.000 | -0.669 | 0.050 | 0.000 | -0.569 | 0.062 | 0.000 | -0.521 | 0.066 | 0.000 | -0.644 | 0.060 | 0.000 |
| Lag = 3 | -0.356 | 0.045 | 0.000 | -0.478 | 0.062 | 0.000 | -0.657 | 0.055 | 0.000 | -0.582 | 0.064 | 0.000 | -0.619 | 0.068 | 0.000 | -0.727 | 0.062 | 0.000 |
| Lag = 4 | -0.178 | 0.041 | 0.000 | -0.360 | 0.060 | 0.000 | -0.605 | 0.058 | 0.000 | -0.524 | 0.063 | 0.000 | -0.623 | 0.074 | 0.000 | -0.745 | 0.065 | 0.000 |
| Lag = 5 | -0.100 | 0.036 | 0.006 | -0.256 | 0.050 | 0.000 | -0.613 | 0.061 | 0.000 | -0.544 | 0.062 | 0.000 | -0.589 | 0.075 | 0.000 | -0.753 | 0.065 | 0.000 |
| Lag = 6 | | | | -0.140 | 0.053 | 0.009 | -0.578 | 0.069 | 0.000 | -0.511 | 0.072 | 0.000 | -0.579 | 0.078 | 0.000 | -0.745 | 0.070 | 0.000 |
| Lag = 7 | | | | -0.110 | 0.052 | 0.037 | -0.573 | 0.072 | 0.000 | -0.465 | 0.081 | 0.000 | -0.480 | 0.087 | 0.000 | -0.715 | 0.081 | 0.000 |
| Lag = 8 | | | | -0.058 | 0.044 | 0.192 | -0.543 | 0.069 | 0.000 | -0.429 | 0.082 | 0.000 | -0.390 | 0.088 | 0.000 | -0.693 | 0.087 | 0.000 |
| Lag = 9 | | | | -0.048 | 0.036 | 0.183 | -0.474 | 0.060 | 0.000 | -0.377 | 0.081 | 0.000 | -0.383 | 0.081 | 0.000 | -0.649 | 0.083 | 0.000 |
| Lag = 10 | | | | -0.034 | 0.030 | 0.256 | -0.339 | 0.045 | 0.000 | -0.284 | 0.081 | 0.001 | -0.271 | 0.082 | 0.002 | -0.445 | 0.081 | 0.000 |
| Lag = 11 | | | | | | | -0.297 | 0.044 | 0.000 | -0.317 | 0.079 | 0.000 | -0.345 | 0.080 | 0.000 | -0.478 | 0.075 | 0.000 |
| Lag = 12 | | | | | | | -0.262 | 0.043 | 0.000 | -0.275 | 0.082 | 0.001 | -0.271 | 0.084 | 0.002 | -0.402 | 0.075 | 0.000 |
| Lag = 13 | | | | | | | -0.185 | 0.036 | 0.000 | -0.169 | 0.079 | 0.035 | -0.034 | 0.085 | 0.689 | -0.198 | 0.075 | 0.010 |
| Lag = 14 | | | | | | | -0.088 | 0.028 | 0.002 | -0.137 | 0.067 | 0.043 | -0.048 | 0.079 | 0.540 | -0.187 | 0.074 | 0.014 |
| Lag = 15 | | | | | | | -0.051 | 0.024 | 0.035 | -0.143 | 0.047 | 0.003 | -0.174 | 0.081 | 0.034 | -0.223 | 0.076 | 0.004 |
| Lag = 16 | | | | | | | | | | -0.121 | 0.045 | 0.009 | -0.095 | 0.083 | 0.256 | -0.156 | 0.075 | 0.042 |
| Lag = 17 | | | | | | | | | | -0.125 | 0.046 | 0.008 | -0.111 | 0.083 | 0.183 | -0.130 | 0.075 | 0.087 |
| Lag = 18 | | | | | | | | | | -0.113 | 0.040 | 0.006 | -0.086 | 0.080 | 0.286 | -0.113 | 0.071 | 0.119 |
| Lag = 19 | | | | | | | | | | -0.068 | 0.030 | 0.026 | -0.039 | 0.069 | 0.577 | -0.148 | 0.070 | 0.039 |
| Lag = 20 | | | | | | | | | | -0.043 | 0.024 | 0.081 | -0.101 | 0.050 | 0.046 | -0.185 | 0.070 | 0.010 |
| Lag = 21 | | | | | | | | | | | | | -0.132 | 0.049 | 0.009 | -0.184 | 0.069 | 0.010 |
| Lag = 22 | | | | | | | | | | | | | -0.064 | 0.049 | 0.201 | -0.123 | 0.069 | 0.078 |
| Lag = 23 | | | | | | | | | | | | | -0.042 | 0.041 | 0.315 | -0.122 | 0.061 | 0.049 |
| Lag = 24 | | | | | | | | | | | | | -0.082 | 0.029 | 0.006 | -0.190 | 0.042 | 0.000 |
| Lag = 25 | | | | | | | | | | | | | -0.075 | 0.024 | 0.002 | -0.188 | 0.041 | 0.000 |
| Lag = 26 | | | | | | | | | | | | | | | | -0.156 | 0.041 | 0.000 |
| Lag = 27 | | | | | | | | | | | | | | | | -0.106 | 0.035 | 0.003 |
| Lag = 28 | | | | | | | | | | | | | | | | -0.086 | 0.025 | 0.001 |
| Lag = 29 | | | | | | | | | | | | | | | | -0.075 | 0.020 | 0.001 |
| **Daily Additional Sick Posts** | | | | | | | | | | | | | | | | | | |
| Lag = 1 | 2.420 | 0.430 | 0.000 | 1.750 | 0.374 | 0.000 | 1.780 | 0.352 | 0.000 | 1.709 | 0.336 | 0.000 | 1.631 | 0.298 | 0.000 | 1.368 | 0.246 | 0.000 |
| Lag = 2 | 2.113 | 0.445 | 0.000 | 1.774 | 0.377 | 0.000 | 1.802 | 0.330 | 0.000 | 1.869 | 0.331 | 0.000 | 1.579 | 0.294 | 0.000 | 1.409 | 0.252 | 0.000 |
| Lag = 3 | 3.296 | 0.472 | 0.000 | 2.732 | 0.398 | 0.000 | 2.637 | 0.339 | 0.000 | 2.554 | 0.343 | 0.000 | 2.126 | 0.313 | 0.000 | 1.928 | 0.265 | 0.000 |

|  | Model 1 (max lag = 5) | | | Model 2 (max lag = 10) | | | Model 3 (max lag = 15) | | | Model 4 (max lag = 20) | | | Model 5 (max lag = 25) | | | Model 6 (max lag = 29) | | |
|---|---|---|---|---|---|---|---|---|---|---|---|---|---|---|---|---|---|---|
|  | *B* | *SE* | *p* | *B* | *SE* | *p* | *B* | *SE* | *p* | *B* | *SE* | *p* | *B* | *SE* | *p* | *B* | *SE* | *p* |
| Lag = 4 | 2.177 | 0.504 | 0.000 | 2.379 | 0.415 | 0.000 | 2.718 | 0.339 | 0.000 | 2.848 | 0.353 | 0.000 | 2.404 | 0.322 | 0.000 | 2.472 | 0.273 | 0.000 |
| Lag = 5 | 2.010 | 0.493 | 0.000 | 2.606 | 0.430 | 0.000 | 2.987 | 0.345 | 0.000 | 2.769 | 0.372 | 0.000 | 2.511 | 0.333 | 0.000 | 2.767 | 0.270 | 0.000 |
| Lag = 6 |  |  |  | 3.611 | 0.454 | 0.000 | 3.546 | 0.358 | 0.000 | 3.347 | 0.360 | 0.000 | 3.151 | 0.322 | 0.000 | 3.285 | 0.259 | 0.000 |
| Lag = 7 |  |  |  | 2.246 | 0.578 | 0.000 | 2.614 | 0.457 | 0.000 | 2.142 | 0.450 | 0.000 | 2.561 | 0.420 | 0.000 | 2.792 | 0.335 | 0.000 |
| Lag = 8 |  |  |  | 1.712 | 0.656 | 0.010 | 2.545 | 0.532 | 0.000 | 1.975 | 0.532 | 0.000 | 2.429 | 0.508 | 0.000 | 2.842 | 0.407 | 0.000 |
| Lag = 9 |  |  |  | 0.962 | 0.634 | 0.132 | 2.661 | 0.555 | 0.000 | 1.889 | 0.565 | 0.001 | 2.392 | 0.557 | 0.000 | 3.021 | 0.454 | 0.000 |
| Lag = 10 |  |  |  | 0.834 | 0.536 | 0.123 | 3.082 | 0.552 | 0.000 | 2.288 | 0.574 | 0.000 | 2.724 | 0.582 | 0.000 | 3.607 | 0.483 | 0.000 |
| Lag = 11 |  |  |  |  |  |  | 3.415 | 0.569 | 0.000 | 3.045 | 0.587 | 0.000 | 3.301 | 0.611 | 0.000 | 4.031 | 0.504 | 0.000 |
| Lag = 12 |  |  |  |  |  |  | 3.639 | 0.577 | 0.000 | 3.537 | 0.584 | 0.000 | 3.813 | 0.573 | 0.000 | 4.978 | 0.493 | 0.000 |
| Lag = 13 |  |  |  |  |  |  | 3.589 | 0.602 | 0.000 | 3.251 | 0.645 | 0.000 | 3.475 | 0.635 | 0.000 | 4.820 | 0.551 | 0.000 |
| Lag = 14 |  |  |  |  |  |  | 3.444 | 0.597 | 0.000 | 2.706 | 0.693 | 0.000 | 2.148 | 0.679 | 0.002 | 4.126 | 0.639 | 0.000 |
| Lag = 15 |  |  |  |  |  |  | 1.993 | 0.495 | 0.000 | 1.132 | 0.671 | 0.096 | 0.459 | 0.667 | 0.494 | 2.465 | 0.652 | 0.000 |
| Lag = 16 |  |  |  |  |  |  |  |  |  | -0.492 | 0.743 | 0.510 | -1.169 | 0.739 | 0.118 | -0.190 | 0.698 | 0.786 |
| Lag = 17 |  |  |  |  |  |  |  |  |  | 0.726 | 0.749 | 0.335 | 1.870 | 0.766 | 0.017 | 2.089 | 0.756 | 0.007 |
| Lag = 18 |  |  |  |  |  |  |  |  |  | 0.409 | 0.775 | 0.599 | 1.071 | 0.821 | 0.196 | 0.716 | 0.729 | 0.330 |
| Lag = 19 |  |  |  |  |  |  |  |  |  | -0.169 | 0.785 | 0.830 | -1.362 | 0.791 | 0.089 | -0.759 | 0.725 | 0.299 |
| Lag = 20 |  |  |  |  |  |  |  |  |  | 0.310 | 0.686 | 0.653 | -0.454 | 0.745 | 0.544 | 0.184 | 0.713 | 0.797 |
| Lag = 21 |  |  |  |  |  |  |  |  |  |  |  |  | 0.497 | 0.762 | 0.517 | 1.001 | 0.727 | 0.174 |
| Lag = 22 |  |  |  |  |  |  |  |  |  |  |  |  | -0.393 | 0.752 | 0.602 | 0.728 | 0.716 | 0.313 |
| Lag = 23 |  |  |  |  |  |  |  |  |  |  |  |  | 0.341 | 0.753 | 0.652 | 0.117 | 0.690 | 0.866 |
| Lag = 24 |  |  |  |  |  |  |  |  |  |  |  |  | -0.215 | 0.742 | 0.773 | -0.238 | 0.638 | 0.710 |
| Lag = 25 |  |  |  |  |  |  |  |  |  |  |  |  | -0.707 | 0.666 | 0.292 | 0.018 | 0.623 | 0.978 |
| Lag = 26 |  |  |  |  |  |  |  |  |  |  |  |  |  |  |  | 0.603 | 0.618 | 0.333 |
| Lag = 27 |  |  |  |  |  |  |  |  |  |  |  |  |  |  |  | 0.193 | 0.632 | 0.761 |
| Lag = 28 |  |  |  |  |  |  |  |  |  |  |  |  |  |  |  | -0.036 | 0.635 | 0.955 |
| Lag = 29 |  |  |  |  |  |  |  |  |  |  |  |  |  |  |  | 0.076 | 0.556 | 0.891 |
|  |  |  |  |  |  |  |  |  |  |  |  |  |  |  |  |  |  |  |
| Adjusted $R^2$ |  |  | 0.900 |  |  | 0.936 |  |  | 0.964 |  |  | 0.970 |  |  | 0.978 |  |  | 0.986 |

**Table E. Model summaries for other COVID-19 posts (difference scores) predicting new cases (difference scores) in mainland China with varying lag terms ($N = 122$)**

| | Model 1 (max lag = 5) | | | Model 2 (max lag = 10) | | | Model 3 (max lag = 15) | | | Model 4 (max lag = 20) | | | Model 5 (max lag = 25) | | | Model 6 (max lag = 29) | | |
|---|---|---|---|---|---|---|---|---|---|---|---|---|---|---|---|---|---|---|
| | *B* | *SE* | *p* | *B* | *SE* | *p* | *B* | *SE* | *p* | *B* | *SE* | *p* | *B* | *SE* | *p* | *B* | *SE* | *p* |
| Intercept | -107.415 | 54.809 | 0.053 | -112.068 | 51.926 | 0.033 | -118.982 | 41.204 | 0.005 | -115.328 | 30.459 | 0.000 | -109.323 | 18.228 | 0.000 | -101.757 | 16.492 | 0.000 |
| Change of Diagnosis Criteria | 12843.360 | 621.189 | 0.000 | 12869.610 | 616.865 | 0.000 | 12845.750 | 528.905 | 0.000 | 11941.980 | 747.882 | 0.000 | 12852.910 | 695.133 | 0.000 | 11878.000 | 810.914 | 0.000 |
| **Daily-Additional New Cases** | | | | | | | | | | | | | | | | | | |
| Lag = 1 | -0.585 | 0.043 | 0.000 | -0.585 | 0.043 | 0.000 | -0.639 | 0.037 | 0.000 | -0.563 | 0.049 | 0.000 | -0.535 | 0.046 | 0.000 | -0.673 | 0.060 | 0.000 |
| Lag = 2 | -0.451 | 0.050 | 0.000 | -0.472 | 0.051 | 0.000 | -0.546 | 0.045 | 0.000 | -0.529 | 0.043 | 0.000 | -0.307 | 0.055 | 0.000 | -0.530 | 0.083 | 0.000 |
| Lag = 3 | -0.320 | 0.053 | 0.000 | -0.371 | 0.055 | 0.000 | -0.492 | 0.049 | 0.000 | -0.511 | 0.038 | 0.000 | -0.172 | 0.064 | 0.009 | -0.436 | 0.100 | 0.000 |
| Lag = 4 | -0.188 | 0.050 | 0.000 | -0.268 | 0.057 | 0.000 | -0.402 | 0.050 | 0.000 | -0.354 | 0.045 | 0.000 | 0.030 | 0.067 | 0.655 | -0.241 | 0.116 | 0.042 |
| Lag = 5 | -0.082 | 0.043 | 0.059 | -0.155 | 0.057 | 0.008 | -0.300 | 0.050 | 0.000 | -0.324 | 0.040 | 0.000 | 0.067 | 0.060 | 0.263 | -0.088 | 0.102 | 0.394 |
| Lag = 6 | | | | -0.078 | 0.057 | 0.174 | -0.205 | 0.049 | 0.000 | -0.281 | 0.039 | 0.000 | 0.034 | 0.053 | 0.521 | 0.008 | 0.077 | 0.922 |
| Lag = 7 | | | | -0.092 | 0.056 | 0.105 | -0.230 | 0.050 | 0.000 | -0.310 | 0.042 | 0.000 | -0.002 | 0.050 | 0.976 | -0.003 | 0.067 | 0.958 |
| Lag = 8 | | | | -0.055 | 0.055 | 0.319 | -0.209 | 0.050 | 0.000 | -0.289 | 0.042 | 0.000 | -0.091 | 0.040 | 0.024 | -0.036 | 0.069 | 0.606 |
| Lag = 9 | | | | -0.063 | 0.050 | 0.207 | -0.235 | 0.049 | 0.000 | -0.315 | 0.041 | 0.000 | -0.187 | 0.032 | 0.000 | -0.081 | 0.064 | 0.209 |
| Lag = 10 | | | | -0.035 | 0.042 | 0.404 | -0.200 | 0.047 | 0.000 | -0.300 | 0.039 | 0.000 | -0.136 | 0.033 | 0.000 | -0.104 | 0.053 | 0.052 |
| Lag = 11 | | | | | | | -0.192 | 0.046 | 0.000 | -0.293 | 0.038 | 0.000 | -0.132 | 0.035 | 0.000 | -0.154 | 0.055 | 0.007 |
| Lag = 12 | | | | | | | -0.194 | 0.045 | 0.000 | -0.311 | 0.037 | 0.000 | -0.099 | 0.038 | 0.012 | -0.154 | 0.043 | 0.001 |
| Lag = 13 | | | | | | | -0.162 | 0.044 | 0.000 | -0.256 | 0.041 | 0.000 | -0.008 | 0.039 | 0.835 | -0.084 | 0.046 | 0.071 |
| Lag = 14 | | | | | | | -0.128 | 0.040 | 0.002 | -0.264 | 0.037 | 0.000 | -0.010 | 0.037 | 0.786 | -0.059 | 0.038 | 0.123 |
| Lag = 15 | | | | | | | -0.092 | 0.034 | 0.008 | -0.252 | 0.034 | 0.000 | -0.056 | 0.034 | 0.110 | -0.064 | 0.037 | 0.093 |
| Lag = 16 | | | | | | | | | | -0.191 | 0.037 | 0.000 | -0.029 | 0.033 | 0.382 | -0.015 | 0.038 | 0.689 |
| Lag = 17 | | | | | | | | | | -0.178 | 0.034 | 0.000 | -0.012 | 0.032 | 0.716 | 0.008 | 0.037 | 0.841 |
| Lag = 18 | | | | | | | | | | -0.170 | 0.032 | 0.000 | -0.035 | 0.032 | 0.273 | -0.028 | 0.036 | 0.448 |
| Lag = 19 | | | | | | | | | | -0.120 | 0.030 | 0.000 | -0.039 | 0.031 | 0.210 | -0.045 | 0.032 | 0.166 |
| Lag = 20 | | | | | | | | | | -0.086 | 0.025 | 0.001 | -0.060 | 0.029 | 0.045 | -0.059 | 0.031 | 0.059 |
| Lag = 21 | | | | | | | | | | | | | -0.021 | 0.034 | 0.537 | -0.044 | 0.032 | 0.174 |
| Lag = 22 | | | | | | | | | | | | | 0.064 | 0.032 | 0.051 | -0.008 | 0.033 | 0.814 |
| Lag = 23 | | | | | | | | | | | | | 0.060 | 0.029 | 0.038 | -0.005 | 0.032 | 0.883 |
| Lag = 24 | | | | | | | | | | | | | 0.050 | 0.024 | 0.041 | 0.012 | 0.033 | 0.716 |
| Lag = 25 | | | | | | | | | | | | | 0.031 | 0.019 | 0.104 | 0.019 | 0.034 | 0.581 |
| Lag = 26 | | | | | | | | | | | | | | | | 0.001 | 0.032 | 0.974 |
| Lag = 27 | | | | | | | | | | | | | | | | -0.009 | 0.032 | 0.784 |
| Lag = 28 | | | | | | | | | | | | | | | | -0.023 | 0.027 | 0.381 |
| Lag = 29 | | | | | | | | | | | | | | | | -0.004 | 0.021 | 0.858 |
| **Daily-Additional Other COVID-19 Posts** | | | | | | | | | | | | | | | | | | |
| Lag = 1 | 0.031 | 0.030 | 0.031 | 0.029 | 0.029 | 0.335 | 0.033 | 0.025 | 0.185 | 0.034 | 0.019 | 0.080 | 0.018 | 0.011 | 0.117 | 0.016 | 0.010 | 0.117 |
| Lag = 2 | 0.044 | 0.030 | 0.044 | 0.060 | 0.029 | 0.044 | 0.069 | 0.025 | 0.007 | 0.046 | 0.019 | 0.019 | 0.043 | 0.011 | 0.000 | 0.039 | 0.010 | 0.000 |
| Lag = 3 | 0.040 | 0.029 | 0.040 | 0.032 | 0.029 | 0.288 | 0.015 | 0.024 | 0.525 | 0.006 | 0.019 | 0.761 | -0.007 | 0.012 | 0.562 | -0.005 | 0.010 | 0.642 |

| | Model 1 (max lag = 5) | | | Model 2 (max lag = 10) | | | Model 3 (max lag = 15) | | | Model 4 (max lag = 20) | | | Model 5 (max lag = 25) | | | Model 6 (max lag = 29) | | |
|---|---|---|---|---|---|---|---|---|---|---|---|---|---|---|---|---|---|---|
| | *B* | *SE* | *p* | *B* | *SE* | *p* | *B* | *SE* | *p* | *B* | *SE* | *p* | *B* | *SE* | *p* | *B* | *SE* | *p* |
| Lag = 4 | 0.044 | 0.031 | 0.044 | 0.050 | 0.032 | 0.119 | 0.040 | 0.026 | 0.121 | 0.043 | 0.022 | 0.052 | 0.041 | 0.013 | 0.003 | 0.039 | 0.011 | 0.001 |
| Lag = 5 | 0.040 | 0.031 | 0.040 | 0.038 | 0.031 | 0.230 | 0.039 | 0.026 | 0.144 | 0.062 | 0.022 | 0.005 | 0.038 | 0.013 | 0.005 | 0.040 | 0.012 | 0.001 |
| Lag = 6 | | | | 0.066 | 0.032 | 0.042 | 0.076 | 0.026 | 0.005 | 0.074 | 0.022 | 0.001 | 0.063 | 0.013 | 0.000 | 0.069 | 0.012 | 0.000 |
| Lag = 7 | | | | 0.022 | 0.032 | 0.500 | 0.058 | 0.027 | 0.035 | 0.053 | 0.022 | 0.020 | 0.031 | 0.014 | 0.033 | 0.046 | 0.013 | 0.001 |
| Lag = 8 | | | | 0.036 | 0.030 | 0.238 | 0.043 | 0.027 | 0.119 | 0.022 | 0.022 | 0.325 | 0.010 | 0.014 | 0.490 | 0.031 | 0.014 | 0.031 |
| Lag = 9 | | | | 0.050 | 0.031 | 0.116 | 0.065 | 0.029 | 0.027 | 0.060 | 0.022 | 0.009 | 0.024 | 0.014 | 0.098 | 0.039 | 0.014 | 0.006 |
| Lag = 10 | | | | 0.094 | 0.031 | 0.003 | 0.090 | 0.028 | 0.002 | 0.088 | 0.022 | 0.000 | 0.051 | 0.014 | 0.001 | 0.053 | 0.013 | 0.000 |
| Lag = 11 | | | | | | | 0.064 | 0.029 | 0.030 | 0.067 | 0.022 | 0.003 | 0.034 | 0.014 | 0.020 | 0.039 | 0.013 | 0.005 |
| Lag = 12 | | | | | | | 0.052 | 0.029 | 0.071 | 0.085 | 0.022 | 0.000 | 0.043 | 0.014 | 0.003 | 0.057 | 0.013 | 0.000 |
| Lag = 13 | | | | | | | 0.077 | 0.028 | 0.007 | 0.099 | 0.023 | 0.000 | 0.078 | 0.014 | 0.000 | 0.096 | 0.013 | 0.000 |
| Lag = 14 | | | | | | | 0.091 | 0.029 | 0.002 | 0.094 | 0.024 | 0.000 | 0.046 | 0.016 | 0.006 | 0.066 | 0.016 | 0.000 |
| Lag = 15 | | | | | | | 0.122 | 0.028 | 0.000 | 0.085 | 0.024 | 0.001 | 0.053 | 0.017 | 0.003 | 0.070 | 0.017 | 0.000 |
| Lag = 16 | | | | | | | | | | 0.005 | 0.025 | 0.851 | -0.059 | 0.017 | 0.001 | -0.043 | 0.017 | 0.017 |
| Lag = 17 | | | | | | | | | | 0.081 | 0.025 | 0.002 | 0.014 | 0.017 | 0.403 | 0.021 | 0.017 | 0.216 |
| Lag = 18 | | | | | | | | | | 0.107 | 0.028 | 0.000 | 0.039 | 0.020 | 0.061 | 0.049 | 0.021 | 0.021 |
| Lag = 19 | | | | | | | | | | 0.000 | 0.042 | 0.995 | -0.006 | 0.030 | 0.842 | -0.024 | 0.038 | 0.520 |
| Lag = 20 | | | | | | | | | | 0.111 | 0.042 | 0.009 | 0.053 | 0.028 | 0.068 | 0.021 | 0.035 | 0.547 |
| Lag = 21 | | | | | | | | | | | | | 0.024 | 0.033 | 0.467 | 0.036 | 0.038 | 0.349 |
| Lag = 22 | | | | | | | | | | | | | -0.068 | 0.037 | 0.072 | 0.004 | 0.039 | 0.927 |
| Lag = 23 | | | | | | | | | | | | | -0.116 | 0.036 | 0.002 | -0.051 | 0.038 | 0.187 |
| Lag = 24 | | | | | | | | | | | | | -0.185 | 0.035 | 0.000 | -0.103 | 0.042 | 0.017 |
| Lag = 25 | | | | | | | | | | | | | -0.156 | 0.035 | 0.000 | -0.105 | 0.035 | 0.003 |
| Lag = 26 | | | | | | | | | | | | | | | | -0.070 | 0.033 | 0.039 |
| Lag = 27 | | | | | | | | | | | | | | | | -0.101 | 0.041 | 0.016 |
| Lag = 28 | | | | | | | | | | | | | | | | -0.052 | 0.053 | 0.327 |
| Lag = 29 | | | | | | | | | | | | | | | | -0.026 | 0.053 | 0.629 |
| | | | | | | | | | | | | | | | | | | |
| Adjusted $R^2$ | | | 0.846 | | | 0.862 | | | 0.913 | | | 0.954 | | | 0.984 | | | 0.988 |

**Table F. Model comparisons for sick posts or other COVID-19 posts (difference scores) predicting new cases (difference scores) in mainland China with varying linear decay rates of the effect of the changed diagnostic criteria on February 12th, 2020 ($N = 122$)**

Hubei province adopted the fifth edition of the diagnostic criteria on Feb. 12th, 2020. We compared models with different decay functions of this change's intervention effects, including an "instant pulse" on Feb. 12th (the indicator was coded as 1 at February 12 and 0 elsewhere), and linear decays in 2 days (the indicator was coded as 1 and 0.5 at February 12th and 13th), 3 days (the indicator was coded as 1, 0.667, and 0.333 at February 12th, 13th, and 14th), 4 days (the indicator was coded as 1, 0.75, 0.5, and 0.25 from February 12th to 15th), or days 5 days (the indicator variable was coded as 1, 0.8, 0.6, 0.4, .2 from February 12th to 16th).

|  | Cases Regressed on Sick Posts | | | | | Cases Regressed on Other COVID-19 Posts | | | | |
| --- | --- | --- | --- | --- | --- | --- | --- | --- | --- | --- |
|  | Adjusted $R^2$ | AIC | BIC | Model df | Residual df | Adjusted $R^2$ | AIC | BIC | Model df | Residual df |
| Instant Pulse | 0.970 | 1741.721 | 1859.490 | 41 | 80 | 0.954 | 1792.342 | 1910.110 | 41 | 80 |
| 2 days | 0.945 | 1815.086 | 1932.855 | 41 | 80 | 0.913 | 1871.580 | 1989.349 | 41 | 80 |
| 3 days | 0.924 | 1854.804 | 1972.573 | 41 | 80 | 0.876 | 1914.615 | 2032.384 | 41 | 80 |
| 4 days | 0.906 | 1881.009 | 1998.778 | 41 | 80 | 0.849 | 1938.742 | 2056.511 | 41 | 80 |
| 5 days | 0.892 | 1897.689 | 2015.458 | 41 | 80 | 0.832 | 1951.363 | 2069.131 | 41 | 80 |

**Table G. Model summaries of sick posts or other COVID-19 posts (difference scores) predicting new cases (difference scores) in mainland China, including a baseline model without effects of social media posts ($N = 122$)**

|  | Model 1 (Baseline) | | | Model 2 (Sick Posts) | | | Model 3 (Other COVID-19 Posts) | | |
|---|---|---|---|---|---|---|---|---|---|
|  | *B* | *SE* | *p* | *B* | *SE* | *p* | *B* | *SE* | *p* |
| Intercept | -106.565 | 55.625 | 0.058 | -99.941 | 22.958 | 0.000 | -115.328 | 28.784 | 0.000 |
| Change of Diagnosis Criteria | 13152.220 | 223.193 | 0.000 | 11476.130 | 682.503 | 0.000 | 11941.980 | 678.476 | 0.000 |
| **Daily Additional New Cases** | | | | | | | | | |
| Lag = 1 | -0.568 | 0.175 | 0.002 | -0.686 | 0.072 | 0.000 | -0.563 | 0.068 | 0.000 |
| Lag = 2 | -0.424 | 0.179 | 0.020 | -0.569 | 0.090 | 0.000 | -0.529 | 0.068 | 0.000 |
| Lag = 3 | -0.302 | 0.162 | 0.066 | -0.582 | 0.082 | 0.000 | -0.511 | 0.069 | 0.000 |
| Lag = 4 | -0.182 | 0.136 | 0.183 | -0.524 | 0.079 | 0.000 | -0.354 | 0.065 | 0.000 |
| Lag = 5 | -0.095 | 0.113 | 0.401 | -0.544 | 0.082 | 0.000 | -0.324 | 0.062 | 0.000 |
| Lag = 6 | -0.031 | 0.088 | 0.728 | -0.511 | 0.102 | 0.000 | -0.281 | 0.058 | 0.000 |
| Lag = 7 | -0.049 | 0.068 | 0.476 | -0.465 | 0.112 | 0.000 | -0.310 | 0.055 | 0.000 |
| Lag = 8 | -0.034 | 0.055 | 0.540 | -0.429 | 0.101 | 0.000 | -0.289 | 0.054 | 0.000 |
| Lag = 9 | -0.082 | 0.045 | 0.070 | -0.377 | 0.099 | 0.000 | -0.315 | 0.053 | 0.000 |
| Lag = 10 | -0.093 | 0.044 | 0.036 | -0.284 | 0.082 | 0.001 | -0.300 | 0.050 | 0.000 |
| Lag = 11 | -0.129 | 0.048 | 0.009 | -0.317 | 0.081 | 0.000 | -0.293 | 0.048 | 0.000 |
| Lag = 12 | -0.148 | 0.057 | 0.011 | -0.275 | 0.078 | 0.001 | -0.311 | 0.049 | 0.000 |
| Lag = 13 | -0.167 | 0.066 | 0.014 | -0.169 | 0.079 | 0.036 | -0.256 | 0.055 | 0.000 |
| Lag = 14 | -0.176 | 0.075 | 0.021 | -0.137 | 0.070 | 0.054 | -0.264 | 0.058 | 0.000 |
| Lag = 15 | -0.182 | 0.081 | 0.028 | -0.143 | 0.055 | 0.011 | -0.252 | 0.056 | 0.000 |
| Lag = 16 | -0.168 | 0.089 | 0.060 | -0.121 | 0.058 | 0.039 | -0.191 | 0.063 | 0.003 |
| Lag = 17 | -0.141 | 0.093 | 0.136 | -0.125 | 0.055 | 0.026 | -0.178 | 0.066 | 0.009 |
| Lag = 18 | -0.139 | 0.098 | 0.161 | -0.113 | 0.053 | 0.037 | -0.170 | 0.070 | 0.017 |
| Lag = 19 | -0.109 | 0.099 | 0.271 | -0.068 | 0.048 | 0.159 | -0.120 | 0.074 | 0.106 |
| Lag = 20 | -0.069 | 0.093 | 0.461 | -0.043 | 0.046 | 0.354 | -0.086 | 0.072 | 0.233 |
| **Daily Additional Posts** | | | | | | | | | |
| Lag = 1 | – | – | – | 1.709 | 0.441 | 0.000 | 0.034 | 0.017 | 0.056 |
| Lag = 2 | – | – | – | 1.869 | 0.428 | 0.000 | 0.046 | 0.024 | 0.060 |
| Lag = 3 | – | – | – | 2.554 | 0.531 | 0.000 | 0.006 | 0.021 | 0.775 |
| Lag = 4 | – | – | – | 2.848 | 0.465 | 0.000 | 0.043 | 0.017 | 0.012 |
| Lag = 5 | – | – | – | 2.769 | 0.443 | 0.000 | 0.062 | 0.025 | 0.014 |
| Lag = 6 | – | – | – | 3.347 | 0.481 | 0.000 | 0.074 | 0.018 | 0.000 |
| Lag = 7 | – | – | – | 2.142 | 0.642 | 0.001 | 0.053 | 0.018 | 0.004 |
| Lag = 8 | – | – | – | 1.975 | 0.687 | 0.005 | 0.022 | 0.019 | 0.236 |
| Lag = 9 | – | – | – | 1.889 | 0.723 | 0.011 | 0.060 | 0.016 | 0.000 |
| Lag = 10 | – | – | – | 2.288 | 0.695 | 0.001 | 0.088 | 0.017 | 0.000 |
| Lag = 11 | – | – | – | 3.045 | 0.793 | 0.000 | 0.067 | 0.019 | 0.001 |
| Lag = 12 | – | – | – | 3.537 | 0.914 | 0.000 | 0.085 | 0.016 | 0.000 |
| Lag = 13 | – | – | – | 3.251 | 0.943 | 0.001 | 0.099 | 0.018 | 0.000 |
| Lag = 14 | – | – | – | 2.706 | 0.889 | 0.003 | 0.094 | 0.019 | 0.000 |
| Lag = 15 | – | – | – | 1.132 | 1.007 | 0.264 | 0.085 | 0.026 | 0.002 |
| Lag = 16 | – | – | – | -0.492 | 0.961 | 0.610 | 0.005 | 0.024 | 0.842 |
| Lag = 17 | – | – | – | 0.726 | 1.017 | 0.477 | 0.081 | 0.020 | 0.000 |
| Lag = 18 | – | – | – | 0.409 | 0.935 | 0.663 | 0.107 | 0.031 | 0.001 |
| Lag = 19 | – | – | – | -0.169 | 0.984 | 0.864 | 0.000 | 0.042 | 0.995 |
| Lag = 20 | – | – | – | 0.310 | 0.800 | 0.700 | 0.111 | 0.046 | 0.017 |
| | | | | | | | | | |
| $R^2$ (Δ) | | | 0.869 | | | 0.980 (0.111) | | | 0.970 (0.101) |
| Adjusted-$R^2$ (Δ) | | | 0.842 | | | 0.970 (0.128) | | | 0.954 (0.112) |
| AIC | | | 1931.403 | | | 1741.721 | | | 1792.342 |
| BIC | | | 1993.092 | | | 1859.490 | | | 1910.110 |

Note. Robust standard errors are reported. $\Delta R^2$ is compared with the baseline model.

**Table H. Model summaries of sick posts or other COVID-19 posts (difference scores) predicting new cases (difference scores) within or outside Hubei ($N = 122$)**

| | Cases Regressed on Sick Posts | | | | | | Cases Regressed on Other COVID-19 Posts | | | | | |
|---|---|---|---|---|---|---|---|---|---|---|---|---|
| | Model 1 (Hubei) | | | Model 2 (outside Hubei) | | | Model 3 (Hubei) | | | Model 4 (Outside Hubei) | | |
| | *B* | *SE* | *p* | *B* | *SE* | *p* | *B* | *SE* | *p* | *B* | *SE* | *p* |
| Intercept | -11.733 | 36.274 | 0.747 | -1.033 | 2.257 | 0.648 | 30.142 | 50.933 | 0.556 | -1.535 | 1.916 | 0.425 |
| Change of Diagnosis Criteria | 13433.280 | 263.711 | 0.000 | -158.715 | 109.008 | 0.149 | 13316.680 | 299.307 | 0.000 | 126.976 | 84.275 | 0.136 |
| **Daily Additional New Cases** | | | | | | | | | | | | |
| Lag = 1 | -0.654 | 0.074 | 0.000 | -0.197 | 0.154 | 0.204 | -0.580 | 0.135 | 0.000 | -0.734 | 0.211 | 0.001 |
| Lag = 2 | -0.577 | 0.087 | 0.000 | -0.379 | 0.191 | 0.051 | -0.475 | 0.145 | 0.002 | -0.331 | 0.190 | 0.084 |
| Lag = 3 | -0.506 | 0.088 | 0.000 | -0.330 | 0.142 | 0.023 | -0.361 | 0.137 | 0.010 | -0.316 | 0.169 | 0.066 |
| Lag = 4 | -0.447 | 0.085 | 0.000 | -0.091 | 0.225 | 0.686 | -0.269 | 0.123 | 0.032 | -0.111 | 0.208 | 0.594 |
| Lag = 5 | -0.408 | 0.084 | 0.000 | -0.245 | 0.184 | 0.188 | -0.201 | 0.110 | 0.072 | 0.264 | 0.230 | 0.254 |
| Lag = 6 | -0.364 | 0.078 | 0.000 | 0.027 | 0.180 | 0.883 | -0.147 | 0.095 | 0.127 | 0.551 | 0.210 | 0.011 |
| Lag = 7 | -0.391 | 0.076 | 0.000 | -0.002 | 0.163 | 0.992 | -0.179 | 0.083 | 0.034 | 0.367 | 0.177 | 0.042 |
| Lag = 8 | -0.385 | 0.070 | 0.000 | -0.213 | 0.160 | 0.188 | -0.181 | 0.074 | 0.016 | 0.129 | 0.155 | 0.407 |
| Lag = 9 | -0.398 | 0.067 | 0.000 | -0.087 | 0.152 | 0.570 | -0.208 | 0.068 | 0.003 | 0.053 | 0.125 | 0.672 |
| Lag = 10 | -0.376 | 0.067 | 0.000 | -0.060 | 0.137 | 0.659 | -0.212 | 0.069 | 0.003 | -0.212 | 0.155 | 0.174 |
| Lag = 11 | -0.360 | 0.069 | 0.000 | -0.036 | 0.107 | 0.740 | -0.230 | 0.072 | 0.002 | -0.325 | 0.174 | 0.065 |
| Lag = 12 | -0.340 | 0.070 | 0.000 | -0.238 | 0.118 | 0.048 | -0.238 | 0.078 | 0.003 | -0.283 | 0.128 | 0.030 |
| Lag = 13 | -0.315 | 0.071 | 0.000 | -0.213 | 0.112 | 0.061 | -0.242 | 0.083 | 0.005 | -0.311 | 0.132 | 0.021 |
| Lag = 14 | -0.293 | 0.072 | 0.000 | -0.015 | 0.167 | 0.931 | -0.233 | 0.087 | 0.009 | -0.178 | 0.106 | 0.095 |
| Lag = 15 | -0.257 | 0.071 | 0.000 | -0.243 | 0.136 | 0.077 | -0.229 | 0.090 | 0.012 | -0.028 | 0.087 | 0.748 |
| Lag = 16 | -0.212 | 0.073 | 0.005 | -0.178 | 0.124 | 0.153 | -0.201 | 0.093 | 0.034 | -0.071 | 0.108 | 0.510 |
| Lag = 17 | -0.166 | 0.076 | 0.032 | -0.159 | 0.109 | 0.149 | -0.158 | 0.096 | 0.101 | 0.197 | 0.177 | 0.269 |
| Lag = 18 | -0.154 | 0.078 | 0.052 | -0.137 | 0.091 | 0.137 | -0.151 | 0.099 | 0.129 | 0.078 | 0.150 | 0.604 |
| Lag = 19 | -0.125 | 0.080 | 0.122 | -0.056 | 0.094 | 0.551 | -0.113 | 0.098 | 0.249 | -0.032 | 0.118 | 0.786 |
| Lag = 20 | -0.073 | 0.078 | 0.350 | -0.111 | 0.103 | 0.286 | -0.068 | 0.091 | 0.457 | -0.030 | 0.091 | 0.738 |
| **Daily Additional Posts** | | | | | | | | | | | | |
| Lag = 1 | 124.065 | 78.630 | 0.119 | 2.649 | 2.489 | 0.290 | 2.915 | 2.722 | 0.287 | 0.155 | 0.045 | 0.001 |
| Lag = 2 | 296.160 | 89.490 | 0.001 | 9.054 | 2.223 | 0.000 | 6.877 | 3.800 | 0.074 | 0.517 | 0.072 | 0.000 |
| Lag = 3 | 359.111 | 108.739 | 0.001 | 6.432 | 2.405 | 0.009 | 5.523 | 3.684 | 0.138 | 0.684 | 0.123 | 0.000 |
| Lag = 4 | 353.444 | 95.466 | 0.000 | 11.500 | 2.657 | 0.000 | 7.933 | 5.480 | 0.152 | 0.505 | 0.156 | 0.002 |
| Lag = 5 | 365.335 | 96.879 | 0.000 | 14.797 | 3.087 | 0.000 | 8.131 | 5.040 | 0.111 | 0.419 | 0.157 | 0.009 |
| Lag = 6 | 369.693 | 95.967 | 0.000 | 14.567 | 3.256 | 0.000 | 5.698 | 4.959 | 0.254 | 0.367 | 0.172 | 0.036 |
| Lag = 7 | 375.696 | 101.524 | 0.000 | 11.298 | 3.035 | 0.000 | 6.754 | 5.029 | 0.183 | 0.344 | 0.198 | 0.086 |
| Lag = 8 | 355.893 | 84.811 | 0.000 | 8.334 | 3.113 | 0.009 | 7.676 | 4.833 | 0.116 | 0.506 | 0.189 | 0.009 |
| Lag = 9 | 370.001 | 80.425 | 0.000 | 4.951 | 2.527 | 0.054 | 8.199 | 5.273 | 0.124 | 0.267 | 0.210 | 0.207 |
| Lag = 10 | 470.718 | 104.710 | 0.000 | 0.691 | 2.334 | 0.768 | 10.591 | 5.241 | 0.047 | -0.425 | 0.211 | 0.047 |
| Lag = 11 | 517.452 | 105.888 | 0.000 | 2.600 | 1.981 | 0.193 | 11.139 | 5.320 | 0.039 | -0.068 | 0.232 | 0.772 |
| Lag = 12 | 500.858 | 116.940 | 0.000 | -2.723 | 2.333 | 0.247 | 13.217 | 6.037 | 0.031 | 0.080 | 0.208 | 0.702 |
| Lag = 13 | 494.104 | 126.973 | 0.000 | 1.487 | 1.767 | 0.403 | 14.840 | 5.959 | 0.015 | -0.020 | 0.141 | 0.888 |
| Lag = 14 | 468.563 | 127.484 | 0.000 | 0.074 | 1.839 | 0.968 | 13.959 | 5.746 | 0.017 | 0.176 | 0.110 | 0.113 |
| Lag = 15 | 475.006 | 118.430 | 0.000 | -2.486 | 2.126 | 0.246 | 15.626 | 5.737 | 0.008 | 0.129 | 0.206 | 0.535 |
| Lag = 16 | 391.729 | 114.233 | 0.001 | -1.549 | 1.616 | 0.341 | 11.981 | 5.475 | 0.032 | 0.041 | 0.192 | 0.832 |
| Lag = 17 | 338.517 | 112.834 | 0.004 | -3.179 | 1.944 | 0.106 | 13.602 | 5.699 | 0.019 | -0.412 | 0.135 | 0.003 |
| Lag = 18 | 270.288 | 97.379 | 0.007 | -3.435 | 1.881 | 0.072 | 10.566 | 5.094 | 0.041 | -0.556 | 0.169 | 0.001 |
| Lag = 19 | 190.939 | 80.451 | 0.020 | 0.726 | 1.716 | 0.674 | 5.955 | 4.644 | 0.203 | 0.094 | 0.212 | 0.658 |
| Lag = 20 | 105.740 | 61.384 | 0.089 | -4.517 | 2.685 | 0.096 | 7.220 | 4.635 | 0.123 | 0.049 | 0.170 | 0.771 |
| $R^2$ (Adjusted $R^2$) | 0.952 (0.928) | | | 0.779 (0.666) | | | 0.911 (0.866) | | | 0.862 (0.790) | | |
| AIC | 1848.179 | | | 1191.557 | | | 1924.113 | | | 1134.491 | | |
| BIC | 1965.948 | | | 1309.326 | | | 2041.882 | | | 1252.259 | | |

Note. Robust standard errors are reported. $\Delta R^2$ is compared with the baseline model.